\documentclass[letterpaper,twocolumn,10pt]{article}
\usepackage{main}
\usepackage{amsfonts,amsmath}       
\usepackage{color}
\usepackage{graphicx}
\usepackage{xcolor}         
\usepackage{colortbl}       

\usepackage{subcaption}
\usepackage{cleveref}
\usepackage{titlesec}

\newcommand{\showFigure}[3]{{
\begin{figure}[!ht]
 \centering
\includegraphics[width=1\linewidth]{#1}
\caption{#2}
\label{#3} 
\end{figure}
}}

\title{Unleashing \textit{Automated} Congestion Control Customization in the Wild}

\begin{document}

\author{\rm Amit Cohen$^1$, Lev Gloukhenki$^1$, Ravid Hadar$^1$, Eden Itah$^1$, Yehuda Shvut$^1$, and Michael Schapira$^{1,2}$\\\rm $^2$Compira Labs, $^2$Hebrew University of Jerusalem}

\maketitle

\begin{abstract}
Congestion control (CC) crucially impacts user experience across Internet services like streaming, gaming, AR/VR, and connected cars. Traditionally, CC algorithm design seeks \emph{universal} control rules that yield high performance across diverse application domains and networks. However, varying service needs and network conditions challenge this approach.

We share operational experience with a system that \textit{automatically customizes} congestion control \textit{logic} to service needs and network conditions. We discuss design, deployment challenges, and solutions, highlighting performance benefits through case studies in streaming, gaming, connected cars, and more.

Our system leverages PCC Vivace, an online-learning-based congestion control protocol developed by researchers. Hence, along with insights from customizing congestion control, we also discuss lessons learned and modifications made to adapt PCC Vivace for real-world deployment.
\end{abstract}

\section{Introduction}

Ensuring high Quality of Experience (QoE) is crucial for Internet services. Yet, despite major infrastructure investments, QoE issues persist in established services like video-on-demand and live streaming.\footnote{Prior studies~\cite{BronzinoSAMTF20} show that, in the US, Netflix and Amazon subscribers with premium-tier Internet access ($\geq 250$Mbps) enjoy video in HD less than $40\%$ of the time on average. This is consistent with our data from multiple content providers/distributors and geographies.} Emerging services such as AR/VR, cloud computing, and connected vehicles pose even greater challenges.

\begin{figure*}
    \centering
    \begin{subfigure}{0.49\textwidth}
        \includegraphics[width=\textwidth]{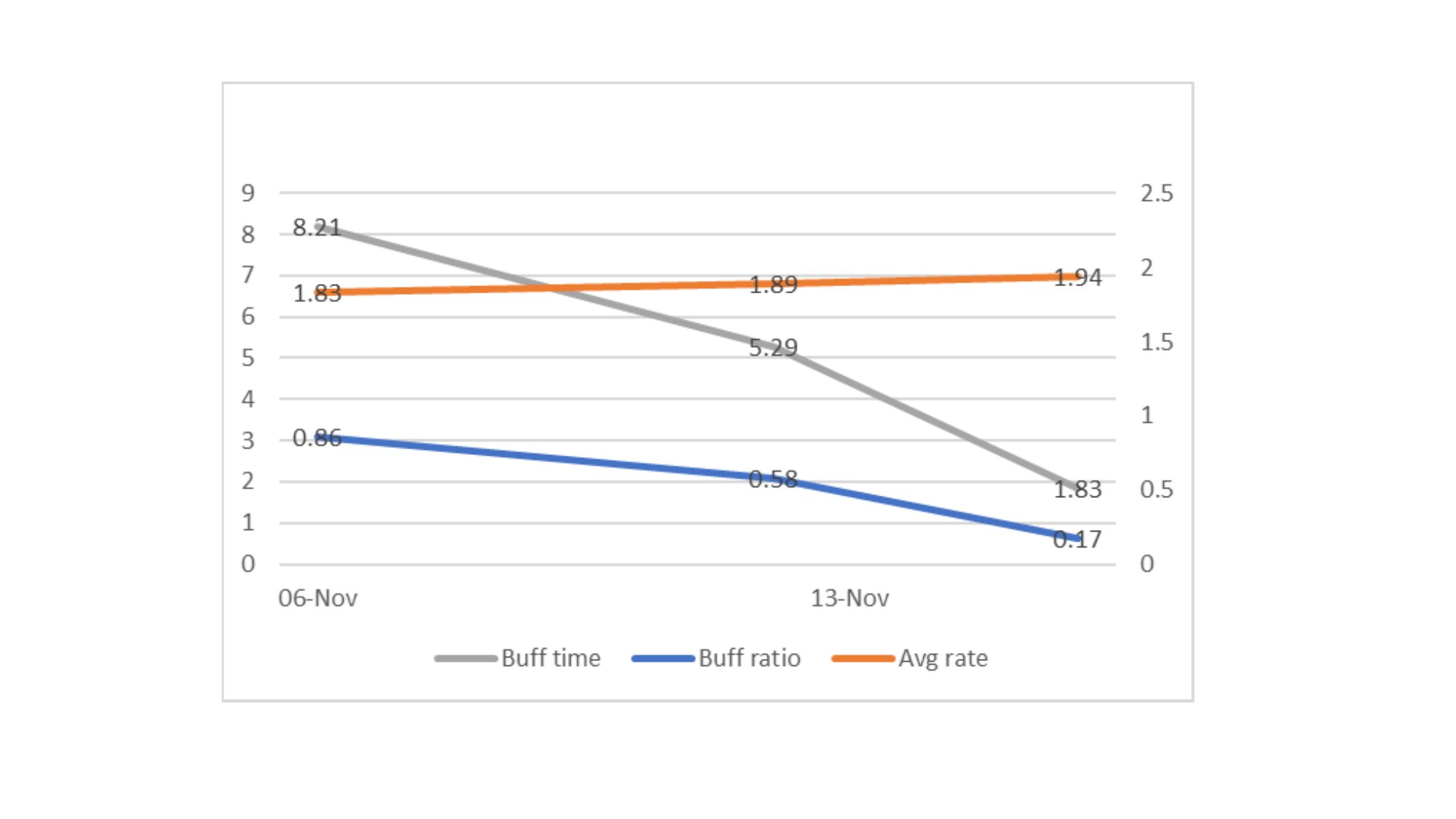} 
        \caption{Customization for a cellular subnet}
        \label{fig:customization-cellular}
    \end{subfigure}
    \begin{subfigure}{0.49\textwidth}
        \includegraphics[width=\textwidth]{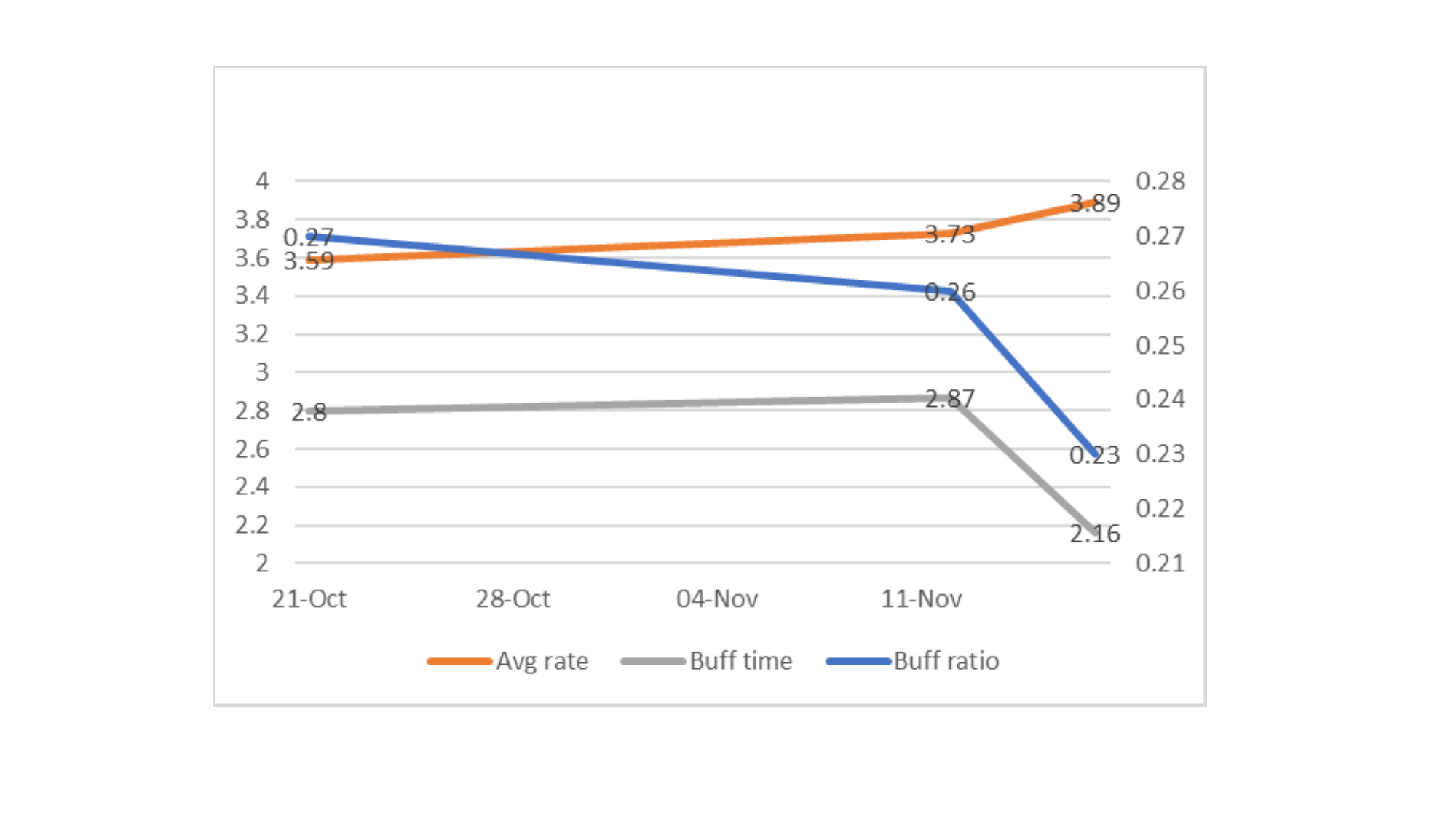} 
        \caption{Customization for a wired subnet}
        \label{fig:customization-wired}
    \end{subfigure}
    \caption{Congestion control customization at work}
    \label{fig:customization@work}
\end{figure*}

User QoE is heavily influenced by \textit{congestion control} (CC), which regulates data flow into the \textit{last-mile} network. Traditionally, CC algorithm design seeks \textit{universal} control rules that provide high performance across different applications and network environments. CC algorithms like TCP Cubic~\cite{CUBIC} and BBR\cite{bbr} reflect this approach. Such \textit{one-size-fits-all} designs ignore service-specific needs (e.g., low latency for live sports \textit{vs.} high throughput for 4K VoD) and are not adapted user-specific conditions (e.g., 5G \textit{vs.} LTE \textit{vs.} wired, high/low competition with cross traffic, etc.). The inherent trade-offs of this one-size-fits-all approach, such as balancing latency and throughput, are well-documented in both empirical~\cite{pantheon} and theoretical studies~\cite{theoretic_cc_bad_sometimes,starvation}.

\autoref{fig:customization@work} shows QoE statistics for live streaming, highlighting how service- and network-specific CC logic adjustments impact performance over time. Each subfigure tracks the impact on QoE of CC logic changes over time for a destination subnet. Three metrics are considered: video bitrate, rebuffering ratio (the fraction of viewing time spent in rebuffering), and absolute rebuffering time (in seconds). The x-axis represents time, and the y-axis QoE performance. Notably, CC adjustments significantly reduced rebuffering without harming—and even slightly improving—video bitrate, demonstrating the benefits of customized congestion control.

As shown in \autoref{fig:customization@work}, customizing CC logic can greatly improve user experience. This, however, is no simple feat. Tuning CC protocols by adjusting parameters (e.g., TCP Cubic’s backoff step or BBR’s probing range) is typically a manual, time-consuming process requiring empirical analysis across various services and geographic locations. As a result, CC logic updates are rare. Recent studies~\cite{configanator,Floo} explore automating CC configuration but focus on simple target performance metrics (e.g., page load times) and mostly discrete parameters.

\vspace{0.1in}\noindent{\bf Operational experience with automated CC customization.} We share insights from designing and deploying a customizable CC system that automatically adapts to service-specific QoE needs and user-specific network conditions. By leveraging numerous CC tuning knobs and \textit{continuous} configuration parameters, our system enables realizing a broad spectrum of CC logic. We detail our adaptation algorithm, which navigates a multi-dimensional parameter space while considering contextual factors (e.g., time-of-day) and ensuring scalability, stability, and safety. We report on six years of operational experience with the system, which has improved QoE across diverse services (including live streaming, short-form videos, VoD, game downloads, and connected cars) and at scale ((>$200$M TikTok videos per server per day, >$20$M game downloads per server per day, many millions of hours of video streaming sessions).

\vspace{0.1in}\noindent{\bf Operational experience with PCC Vivace.} Operational Experience with PCC Vivace. A key part of our solution is a remotely configurable variant of Performance-oriented Congestion Control (PCC)\cite{pcc,vivace}, namely, PCC Vivace~\cite{pcc,vivace,pcc_proteus,aurora}. We share our experience transitioning this research prototype to production, including the challenges faced in making it field-ready.


\section{System Overview}

\showFigure{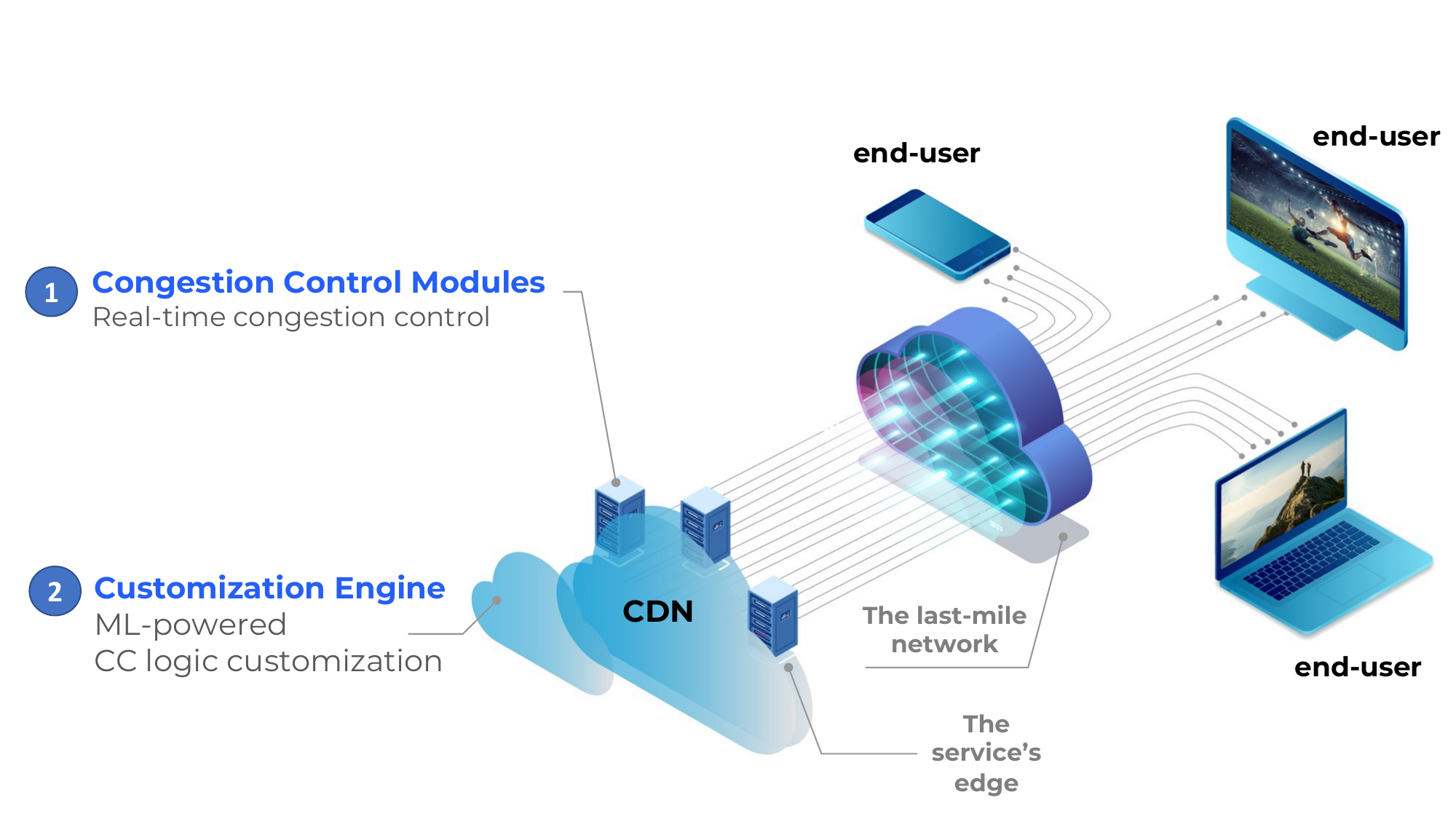}{High-level architecture.}{fig:CDN-solution}

\autoref{fig:CDN-solution} illustrates the architecture of our customizable congestion control solution, using data delivery from a CDN's edge as an example. The architecture consists of two components:

\vspace{0.1in}\noindent{\bf Component I: Congestion control modules at the service edge.}. The edge of the service, where the content resides, varies across services: CDN nodes / video caches for video streaming, end-user devices / servers for videoconferencing, edge servers for cloud gaming, connected cars, etc. The CC modules are remotely configurable through an API, allowing adjustments to the CC logic for controlling traffic transmission rates.

\vspace{0.1in}\vspace{0.1in}\noindent{\bf Component II: Centralized, cloud-based customization engine.} This engine adapts the CC logic of the CC modules based on service-level requirements and experienced network conditions for different destinations. It continuously collects statistics from the CC modules, and possibly the service providers/distributors, to periodically reconfigure the CC logic.

\vspace{0.1in}These two components implement different functionalities (online rate optimization \emph{vs.} CC logic customization), operate at different timescales (milliseconds \emph{vs.} $10$s of minutes), use different types of data (transport-layer only \emph{vs.} possibly also application-layer), target different performance metrics, apply different algorithms, and entail different operational challenges. The division of labor between the two components reflects key design principles: 
\begin{itemize}
    \item \textbf{Robust edge \textit{vs.} statistical learning in the cloud.} CC at edge is expected to be robust even under \textit{adversarial} conditions, with adaptation environmental regularities delegated to the cloud.
    \item \textbf{Avoiding cloud-edge fate sharing.} Cloud failures should not impact real-time performance; the edge continues with its last configuration until the cloud is restored.
    \item \textbf{Limited edge state.} The edge relies only on local, online, transport-layer data, with cross-layer optimization and deeper processing handled offline in the cloud.
\end{itemize}

\subsection{A Configurable PCC Vivace}\label{subsec:parametrized-PCC}

PCC builds on ideas and optimization techniques from online learning theory and game theory. PCC Vivace's online rate optimization consists of a slow-start phase, after which Vivace executes online gradient ascent on a \textit{utility function} of the form
$u(x)=\alpha x^\delta-\beta xL-\gamma x\frac{dRTT}{dt}$, where $x$ is the sending rate, $L$ is the experienced packet loss rate, $\frac{dRTT}{dt}$ is the rate of RTT change, and $\alpha,\beta,\gamma\geq 0$, and $0<\delta<1$ are constants. The step size $\omega>0$ governs the gradient update. Our CC modules employ \emph{parametrized} PCC Vivace logic, with configurable parameters including $\alpha,\beta,\gamma,\delta$, and $\omega$, as well as other natural parameters: the initial sending rate, latency and loss thresholds for exiting slow-start, maximum permissible rate during slow-start, filtering levels for latency and loss (where latency/loss measurements under these levels are ignored), and more.

We chose PCC Vivace as the CC logic for several reasons: 

\vspace{0.1in}\vspace{0.1in}\noindent{\bf PCC Vivace's logic is highly amenable to configurability.} The selected parameters significantly impact protocol behavior and have clear semantic meanings, making it easier to reason about and debug. Different parameter assignments lead to distinct CC behaviors. For example, higher values of $\beta$ and $\gamma$
(loss and delay penalties) make Vivace more sensitive to loss and latency, respectively, while higher values of $\alpha$ and $\delta$ make the protocol more throughput-hungry.

\vspace{0.1in}\vspace{0.1in}\noindent{\bf Parametrized PCC logic spans a broad spectrum of CC behaviors.} Our implementation allows configuration of multiple parameters from continuous ranges, enabling a wide diversity of CC logic. This offers greater flexibility than simply choosing from a limited set of existing CC protocols, or from a discrete set of protocol configuration options, which can be seen as points within this multidimensional space.

\vspace{0.1in}\vspace{0.1in}\noindent{\bf PCC Vivace provides useful, \textit{provable} guarantees.} Vivace offers provable guarantees, including bounded worst-case regret for individual senders and convergence to a fair rate equilibrium for multiple senders~\cite{vivace}. These guarantees pertain to \textit{ranges} of values for PCC parameters (e.g., all $\alpha,\beta,\gamma>0$, and all $0<\delta<1$).

\subsection{The E2E Customization Process}\label{subsec:e2e-customization-process}

To illustrate the end-to-end customization process, we focus on a single PCC module, a specific traffic type (e.g., live sports), and a group of destinations (e.g., a cellular subnet within a certain ISP). Let $\theta$ be the set of configurable parameters of the PCC module (see \autoref{subsec:parametrized-PCC}). At intervals of $\Delta t$, the cloud selects value assignments for $\theta$. After each assignment $\theta_t$ for time interval $t$, the cloud observes performance-related statistics from that time interval and applies a \emph{reward function} to compute a real-valued \emph{reward} $r_t$. Using historical statistics, rewards, and additional context (e.g., time-of-day), the cloud selects a new configuration $\theta_{t+1}$, and so forth. The cloud's objective is to maximize the average reward across time, i.e., $\frac{\Sigma_{t=1}^T r_t}{T}$.

We illustrate this process with a simple example: a PCC module installed on a CDN edge node serving game download requests (e.g., PlayStation, Xbox) from clients. Consider an IP subnet $X$, where traffic to destinations in $X$ shows similar trends. The cloud periodically (say, every $20$ minutes) updates the PCC configuration $\theta$ for connections with destinations in $X$. After the inter-configuration time interval ($20$ minutes) elapses, the cloud aggregates transport-layer and/or application-layer statistics into a performance score. Consider a simple reward function of the form $r_t=c_{T} T - c_L L$. $T$ here represents the average game download throughput (total size in bytes divided by total download duration). $L$ represent the average packet loss rate across game downloads. $c_{T}$ and $c_L$ are constant coefficients. $T$ can be derived from application-layer CDN logs (e.g., NGINX logs or QUIC logs), whereas $L$ can be derived from correlating transport-layer statistics with CDN logs. This reward can be extended to include additional factors (\textit{e.g.}, RTT) or different percentiles of performance metrics. Engineering the reward to align with desired performance is crucial, as shall discussed below.

\section{A Closer Look at Customization}

We next elaborate on four important aspects of the cloud-based customization process: (1) How often are configurations changed? (2) At what aggregate granularity is this done? (3) What is the input to the customization process? (4) What is the target objective?

\subsection{Customization Granularity}

Customizing CC involves a tradeoff between specificity (e.g., optimizing PCC for each destination IP) and the need for statistically meaningful data, which requires enough data points. Similarly, there is a tradeoff between making timely decisions (frequent CC logic changes) and waiting long enough to gather reliable statistics. Below, we discuss our choices for configuration-aggregate and configuration-time granularity, which empirically balance informed decision-making with timeliness.

\vspace{0.1in}\noindent{\bf Configuration-aggregate granularity.} In our framework, the basic decision-making granularity is \textit{(service type, destination IP subnet)}. Grouping by service type is important due to varying QoE requirements across services. Service type is typically known to us in advance (e.g., when installing on a CDN node for TV or TikTok content). Grouping by IP subnet allows for sufficient data collection within short time periods and suggests similar network conditions across destinations. We typically use $8-20$ IP subnet aggregates per server (as supported by the Linux kernel). Dividing the IP destination space to clusters in a manner that both yields fairly homogeneous clusters in terms of network conditions and avoids skewed traffic distribution is an interesting algorithmic challenge. See \autoref{sec:discussion} for discussion.

\vspace{0.1in}\noindent{\bf Configuration-time granularity.} Our empirical experience shows that adapting CC logic every $10-20$ minutes strikes a balance between timely and informed decisions. The granularity should ensure that statistics for similar time periods (e.g., same day, same time-of-day) do not vary significantly. Different content types (e.g., short-form videos, game downloads) generate varying traffic volumes, requiring adjustments to the configuration-time granularity, a process that can be automated.

\subsection{Data Sources for Customization}\label{subsec:customization-detailed}

The customization engine can utilize two sources of data: (1) aggregate transport-layer statistics gathered by the PCC modules, and, potentially, also (2) application-layer statistics, provided by the content provider/distributor.

In our implementation, each active transport-layer connection sends an aggregate data point every second, including statistics such as average throughput, minimum/average/maximum PCC sending rate, packet loss rate, minimum/average RTT, time in slow-start, and app-limited time. Provided application-layer statistics might include CDN logs and QoE measurements (e.g., video rebuffering and quality).

\subsection{\bf Reward Engineering}\label{subsec:reward-engineering}

Our cloud adapts each PCC module's configuration to optimize the relevant performance metric, which varies by service. For video streaming, the metric may combine video quality and rebuffering time; for game downloads, it could be download time; for real-time apps, it may reflect latency. Different content distributors may quantify QoE differently even for the same content type. Reward engineering involves selecting the relevant metric from the service's perspective, either based on predefined metrics from the distributor or through joint discussions.

We next discuss two important aspects of reward engineering: (1) \textit{explicitly} capturing desired tradeoffs, and (2) identifying the relevant data sources.

\vspace{0.1in}\noindent{\bf Explicitly capturing desired performance tradeoffs.} Optimizing QoE is often a multiobjective problem, where improving one metric can degrade another. For example, lower latency in real-time video applications may reduce video quality to avoid exceeding network capacity. A reward function must explicitly capture these tradeoffs. For instance, consider the game download reward function in  \autoref{subsec:e2e-customization-process} $r_t=c_{T} T - c_L L$, where $T$ and $L$ are the average game throughput and data loss respectively.$T$ and $L$ can be normalized with factors $\tau_T$ and $\tau_L$ such that $1$ unit of $T$ corresponds to $1$Mbps and $1$ unit of $L$ corresponds to $1\%$ loss. Choosing coefficients $c_{T}=\tau_T$ and $c_{L}=10\tau_L$ sets the tradeoff such that increasing throughput by $1$Mbps is deemed harmful if the loss rate exceeds $0.1\%$ and beneficial otherwise. More complex tradeoffs, like penalizing certain percentiles or when a threshold is exceeded, can also be captured.

\vspace{0.1in}\noindent{\bf Surrogate reward functions.} For effective customization, performance data used to compute rewards must align with the reconfiguration update rate $\Delta t$. While transport-layer data is constantly available to us, which is provided by the content distributor, might not be. For example, a third-party CDN distributing TikTok content has continuous access to its own logs but may receive TikTok’s QoE statistics (e.g., video stall frequency) at a slower rate (e.g., only daily digests). Similarly, CDNs relying on third-party platforms like Conviva or NPAW for video QoE scores face the same issue. In such cases, we use ``\textit{surrogate reward functions}'', which rely on readily available data that is strongly correlated with the target QoE metrics. For example, in HTTP-based video streaming, download throughput correlates with video bitrate, and the total download time of initial HTTP responses correlates with video start time. A surrogate reward function could use CDN log data (e.g., download sizes and durations) to optimize the correlated QoE metrics, which are typically measured at the client.

\vspace{0.1in}\noindent{\bf Scalable reward engineering.} Importantly, which reward engineering is inherently manual and service-specific, it is not tailored to network, geography, or device type. In addition, insights gained from reward engineering for the same content type often transfer across content distributors. As a result, reward engineering is infrequent, typically occurring only when applying the solution to new content or new distributors, and builds on past operational experience.

\section{CC Customization Algorithms}\label{sec:customization-algorithm}

In this section, we present the customization algorithm employed by the cloud element of our solution. We start by providing intuition for our approach. We then present the basic algorithm and discuss various enhancements.

\subsection{\bf Intuition: Regret Minimization}\label{sec:warmup}

Consider a simple scenario where a sender repeatedly transmits a fixed-size file over a single link with static network parameters (bandwidth, latency, buffer size, and non-congestion-induced loss). The sender selects a CC protocol (e.g., TCP Cubic, BBR, PCC Vivace, Copa) for each transfer, aiming to maximize the average reward across transmissions. The reward for a single transfer captures the tradeoff between download speed and packet loss, and is defined as $x-cxL$, where $x$ is the file transfer duration, $L$ is the packet loss rate, and $c$ is a constant coefficient. Rewards are noramized to the $[-1,1]$ interval.

This challenge naturally maps to a \textit{multi-armed bandit} (MAB) problem \cite{multi_arm_bandits,hazanbook}. In MAB, an \textit{agent} repeatedly selects an action $a_t$ from an action set $A$ at each time $t$ and observes a reward $r_t$. The agent's goal is to maximize the average reward $\frac{\Sigma_t=1^T r_t}{t}$ for some large $T>0$. We consider the \textit{probabilistic} MAB setting, where each action's reward is a random variable. MAB algorithms aim to minimize \textit{regret}--the gap between the best fixed action in hindsight and the agent’s actual performance. See \cite{multi_arm_bandits,hazanbook} for details.

In our context, the agent is the traffic sender, time steps correspond to file transfers, actions are CC protocol choices, and the reward is $x-cxL$. Reward values are stochastic due to both external factors (e.g., non-congestion loss) and protocol behavior. A regret-minimizing MAB algorithm ensures the sender’s average reward approaches that of the best fixed CC protocol in hindsight.

The above discussion illustrates how \emph{configurable} CC logic can \emph{automatically} adapt to optimize performance. Next, we extend this intuition to more realistic environments.

\subsection{Contextual Continuum-Armed Bandits}\label{subsec:CMAB}

To handle real-world CC customization, we enhance the simple MAB framework to CC logic adaptation in \autoref{sec:warmup} in two key ways: (1) \textbf{Context-aware decisions}. Network conditions follow temporal patterns (\textit{e.g.}, diurnal cycles, special events). We incorporate \emph{context}, such as ``midnight-to-8AM on weekdays'' or more granular insights, into CC adaptation. (2) {\bf Continuous action space.} Since CC parameters vary continuously, decision-making must account for a \textit{continuum} of possible actions.

Formally, we extend the MAB setting in \autoref{sec:warmup} as follows. Before selecting an action, the agent observes a context $c_t$ from a fixed discrete set $C$. The reward $r_t$ is derived from a context-specific reward function $R_c$, which can vary across contexts. The goal remains regret minimization—minimizing the gap between the achieved average reward and the best fixed action \textit{per context} in hindsight. For details on MAB and its contextual and continuum-armed extensions, see~\cite{hazanbook,multi_arm_bandits}.

We employ a contextual continuum-armed bandit algorithm, optimizing the PCC configuration for each context $c\in C$ separately. After observing context $c_t\in C$ at time $t$, the algorithm updates the PCC configuration for context $c=c_t$, $\theta_{c,t}$, via a gradient step with respect to the reward function. The gradient is estimated via a standard ``trick''~\cite{hazanbook}: sampling a random point ``around'' the current PCC configuration and using the observed reward $r_t$ to derive an unbiased estimator for the reward gradient $g_t$. The process iterates as follows: 

\begin{enumerate}
    \item Observe context $c_t\in C$.
    \item Sample $x_t\mathbb(S_1)$ at random, where $x_t$ is a $|\theta|$-dimensional vector on the unit-sphere $\mathbb{S}_1$.
    \item Set $y_t=\theta_{c_t,t}+\phi x_t$, for a fixed small constant $\phi>0$.
    \item Set the PCC configuration to be $y_t$ and observe reward $r_t$.
    \item Compute $g_t=r_t x_t$
    \item Set $\theta_{c,t+1}=\theta_{c,t}+\eta g_t$.
\end{enumerate}
  
\subsection{Practical Considerations}

\vspace{0.1in}\noindent{\bf Scalability.} Given unlimited time, the algorithm in \autoref{subsec:CMAB} converges to high-performing configurations. However, scalability may be hindered by context proliferation, where infrequent contexts receive fewer updates, slowing learning. Additionally, high-dimensional parameter spaces complicate accurate reward gradient estimation.

In practice, setting the default context set $C$ to six contexts based on time-of-day and day-of-week, e.g., $\{16:00-00:00,00:00-8:00,8:00-16:00\}\times\{weekday,weekend\}$ already yields significant improvements. To manage configurable PCC parameters, our implementation periodically (every few weeks) and automatically evaluates each parameter's correlation with observed rewards by testing individual assignments while keeping others fixed. The most impactful ($3-4$) parameters are then set as configurable, with the rest fixed. Only parameters with potential for improvement are chosen, ensuring gradual exploration of previously unoptimized parameters.

\vspace{0.1in}\noindent{\bf Stability.} The convergence of our gradient ascent algorithm is affected by two types of noise: (1) noisy network measurements and (2) gradient estimation noise due to random sampling (see \autoref{subsec:CMAB}). To address this, we incorporate two measures: (1) smoothing convergence by updating the gradient using a weighted average of the current and previous gradients (momentum), and (2) reducing noise by averaging multiple gradient estimates from different samples ($x_t$'s).

\vspace{0.1in}\noindent{\bf Safety.} If the network changes radically (\textit{e.g.}, due to outages, routing changes), a PCC configuration learned in the ``old'' environment may perform poorly. While the customization engine will eventually adapt to the new environment, this may take too long. To address this, our algorithms include defaulting rules that revert to a ``safe configuration'' if rewards drop persistently below the average or if packet loss and latency are uncharacteristically high. This safe configuration is expected to provide robustly acceptable performance, and is informed by operational experience in many environments and prior experiments. The safe configuration then serves as the new starting point for the continuous customization process.

\section{PCC Vivace: From Lab to Field}\label{sec:prototype-to-real}

Our implementation of the PCC Vivace module is based on existing research prototypes (UDT library~\cite{vivace} and Linux~\cite{jay18pcc-kernel}). Academic evaluations of PCC Vivace typically focus on environments with elastic traffic, high Bandwidth-Delay products, and limited changes to network conditions. However, in real-world networks, our initial implementation showed erratic behaviors due to deviations from these conditions. To address this, we made various modifications to Vivace. We now discuss these changes.

\subsection{Implementation and Deployability}\label{subsec:edge-implementation}

\vspace{0.1in}\noindent{\bf Linux kernel and QUIC implementations.} Our PCC Vivace module is implemented as both a Linux kernel module and in LiteSpeed QUIC, incorporating modifications to the base PCC Vivace algorithm. These implementations support remote configuration of PCC parameters. The Linux kernel version leverages Google's BBR mechanisms for pacing and transport-layer statistics. As noted in~\cite{vivace}, like BBR, the sender-side, TCP-compatible Linux kernel module allows easy deployment with no receiver-side changes. The module can be installed as a ``hot update'' without interrupting ongoing connections or recompiling the kernel.

\vspace{0.1in}\noindent{\bf The thin agent.} Our Linux implementation includes a so called``thin agent'' that aggregates TCP socket statistics and relays them to the cloud. It also forwards reconfiguration commands from the cloud to the PCC module. In the QUIC implementation, the thin agent is integrated into the all-user-space QUIC code.

\subsection{\bf Accelerating Vivace's Decision Making}

PCC Vivace's decision-making occurs at the granularity of \textit{monitor intervals} (MIs), typically lasting $2$ RTTs. Rate adaptation involves $2-4$ MIs for gradient estimation, followed by a rate update. This leads to spending too much time exploring rates and too little time exploiting good rate choices, and to reacting slowly to network changes. To address these issues, our PCC modules incorporate the following modifications to PCC Vivace.

\vspace{0.1in}\noindent{\bf Unifying exploration and exploitation.} In our implementation, exploration and exploitation are unified. When transitioning from rate $r_1$ to $r_2 = r_1 + \epsilon$, the utility derivative at $r_2$ is approximated by estimating the gradient at $\frac{r_1 + r_2}{2}$, using $\xi = \frac{r_2 - r_1}{\epsilon}$. This approach accelerates decision-making by $3-5$x.

\vspace{0.1in}\noindent{\bf Early decision-making.} When a sender receives many ACKs (e.g., $100$ per RTT), our implementation allows early decisions after receiving enough ACKs (e.g., $50$) without waiting for the full $2$ RTTs to elapse. Importantly, however, only early \textit{rate decreases} are permitted, as they benefit both the sender, whose negative utility gradient indicates that a lower rate would improve its performance, and competing traffic, by preventing fast senders from monopolizing bandwidth.

\subsection{Reacting to Hazardous Conditions}

Unlike TCP and BBR, PCC Vivace never re-enters slow-start and relies solely on the gradient ascent process after leaving slow-start. However, this can be problematic under sudden network deterioration. To address this, we introduce ``\textit{emergency breaks}'' that significantly reduce the rate in extreme conditions. These breaks are triggered by (1) high packet loss (e.g., $>10\%$), (2) excessive latency, or (3) a large gap between sending rate and measured throughput (when traffic is not app-limited). The thresholds are set to ensure emergency breaks are rare and only triggered under severe conditions.

\subsection{Dealing with Inaccurate Gradients}

\vspace{0.1in}\noindent{\bf Contending with insufficient ACKs.} Accurate gradient estimation can be haed when too few packets are sent in an MI, such as with slow sending rates, app-limited traffic, or low RTTs (e.g., $1$ms in 5G networks). To address this, our implementation (1) flags MIs as app-limited and ignores them if too few packets are sent, and (2) prolongs MIs until enough packets are ACKed when the BDP is low.

\vspace{0.1in}\noindent{\bf Filtering noise.} Noisy statistics, from delayed kernel batching or measurement errors, can distort utility gradient predictions. To mitigate this, our implementations filter out outliers and ensure that observed phenomena, such as packet loss, meet persistence criteria (\textit{e.g.}, only consider packet losses when exceeding a threshold and if including non-consecutive packet losses).

\subsection{Safety and Fairness via Constrained PCC Configuration} 

To avoid exploring PCC configurations that induce unacceptable QoE, or inflict unacceptable collateral damage, the PCC configurations outputted by the cloud are constrained. Here, the clear semantic meaning of the PCC parameters is useful. For instance, to avoid suffering high packet loss, and be fair towards cross traffic, the penalty for loss in PCC's utility function is lower bounded. Our space of permissible PCC configurations was informed by controlled experiments. See examples of such experiments, establishing PCC Vivace's friendliness towards legacy TCP, in~\cite{vivace}.

\begin{figure*}[h!]
    \centering
    \begin{subfigure}{0.49\textwidth}
        \includegraphics[width=\textwidth]{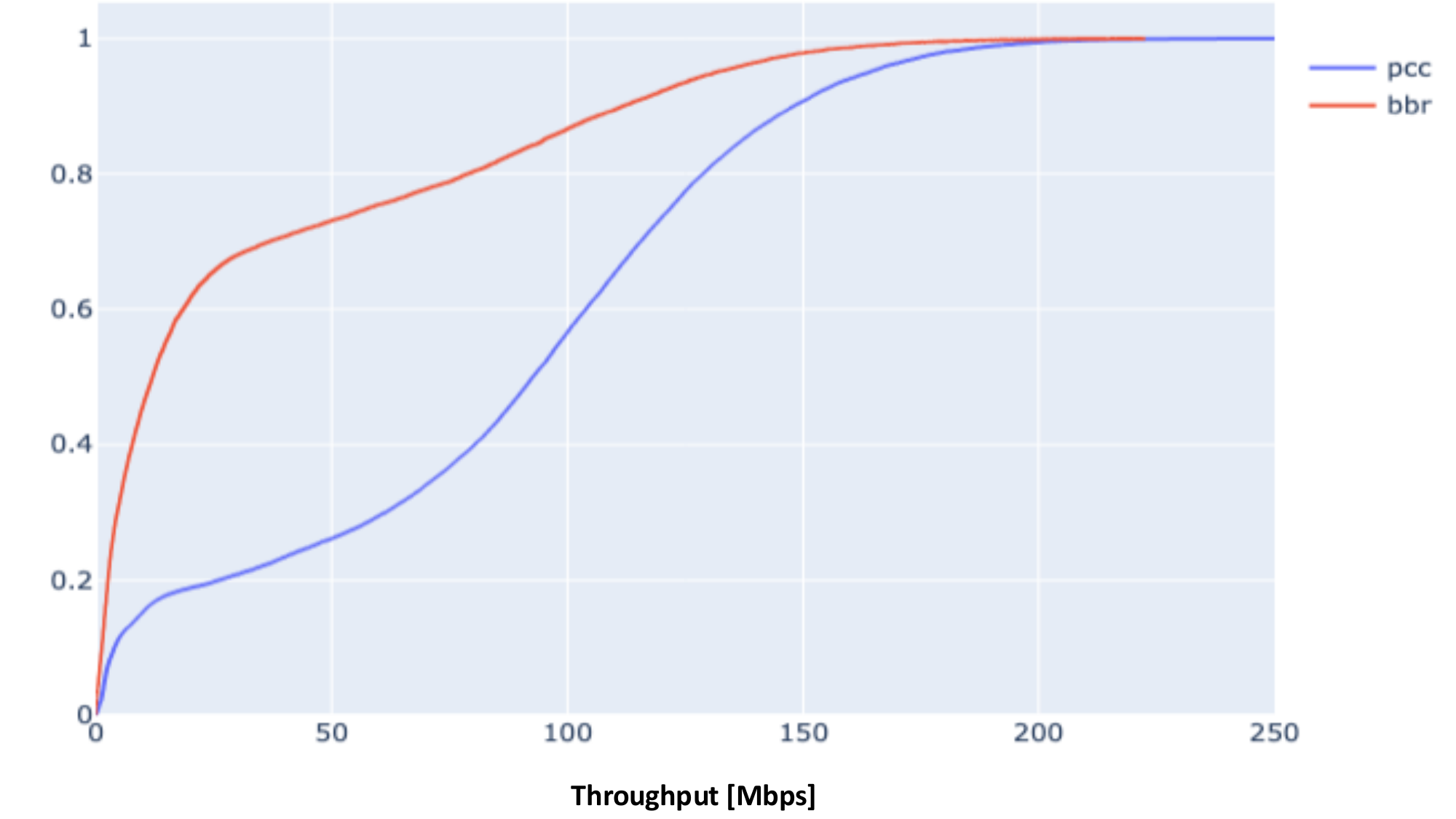} 
        \caption{CDF of throughput}
        \label{fig:TikTok-tpt}
    \end{subfigure}
    \begin{subfigure}{0.49\textwidth}
        \includegraphics[width=\textwidth]{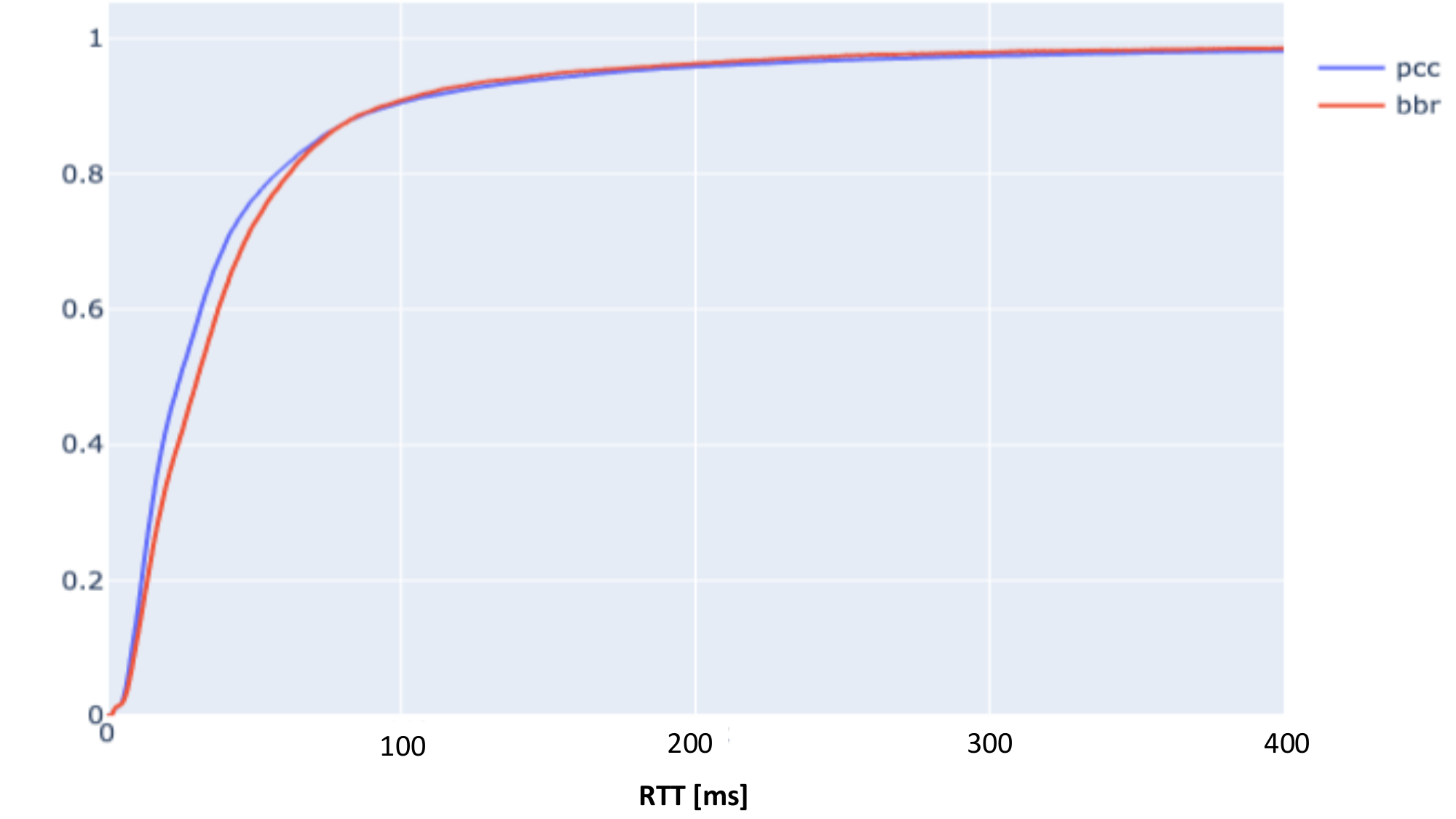} 
        \caption{CDF of RTT}
        \label{fig:TikTok-RTT}
    \end{subfigure}
    \caption{Results from live deployment for TikTok content}
    \label{fig:TikTok-rewards}
\end{figure*}

\showFigure{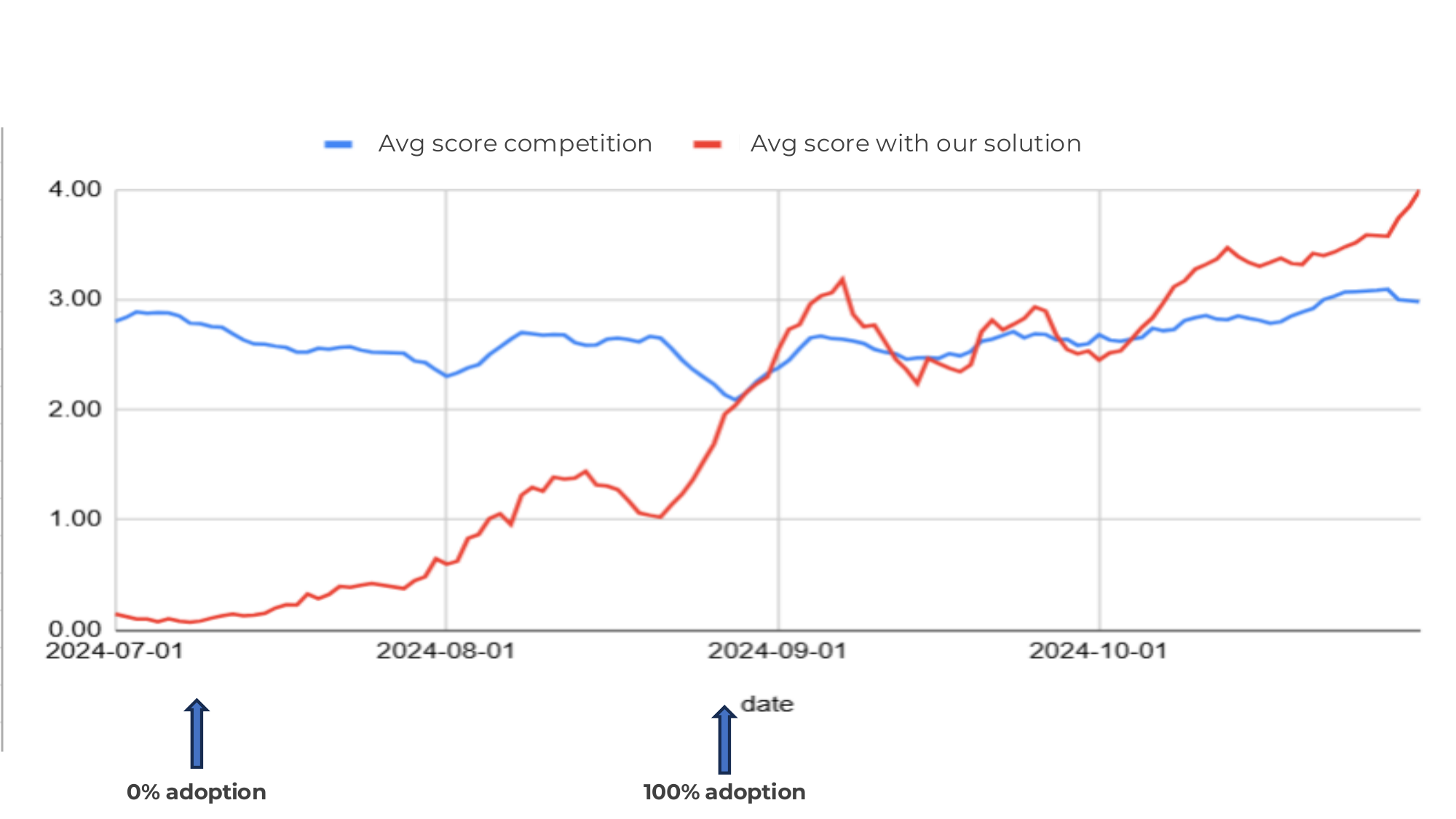}{Implications for TikTok user experience.}{fig:TikTok-QoE}

\showFigure{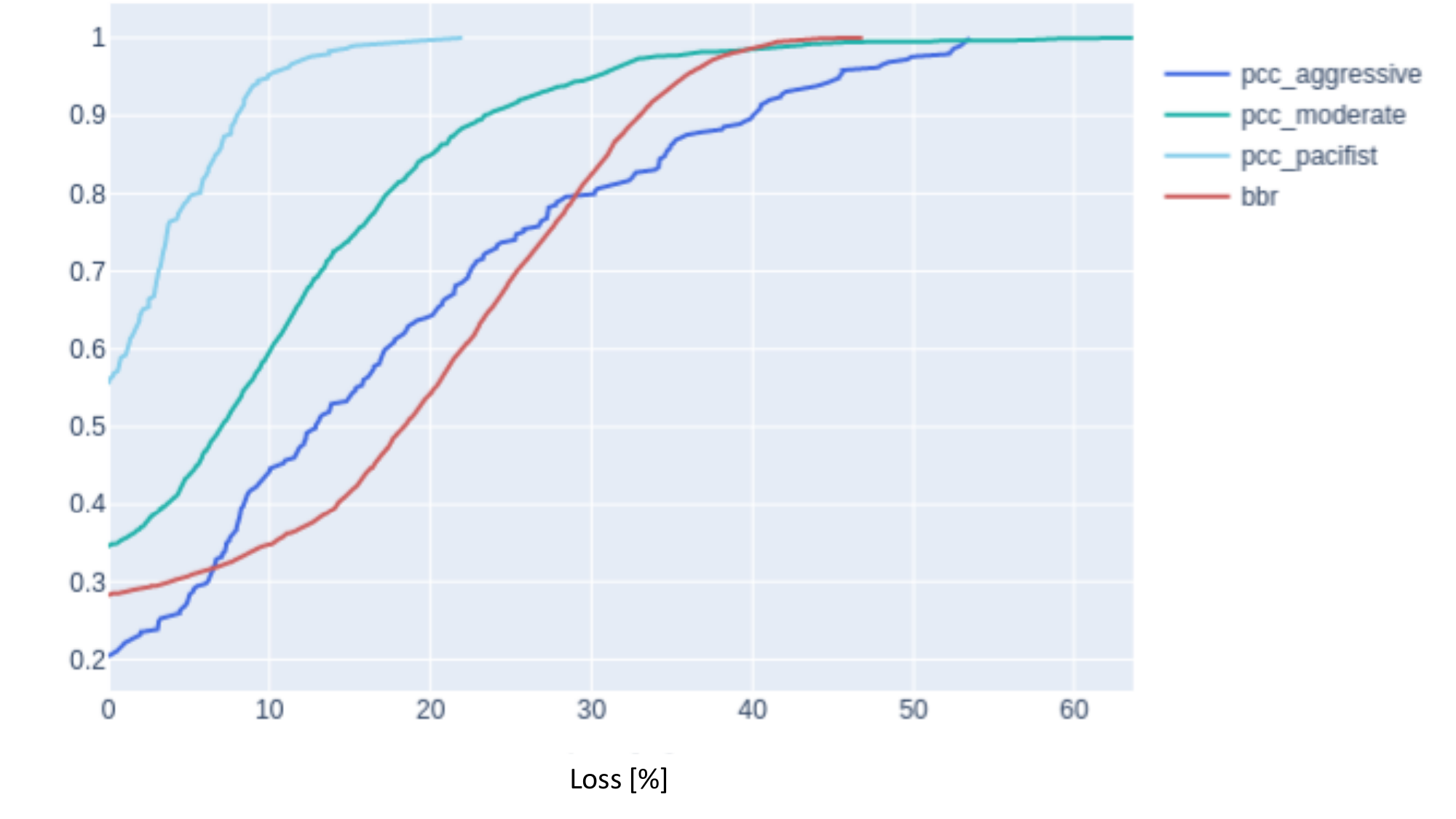}{Loss comparison of BBR and the $3$ learned PCC configurations}{fig:meta-loss}

\begin{figure*}[h!]
    \centering
    \begin{subfigure}[t]{0.49\textwidth}
        \centering
        \includegraphics[width=\textwidth]{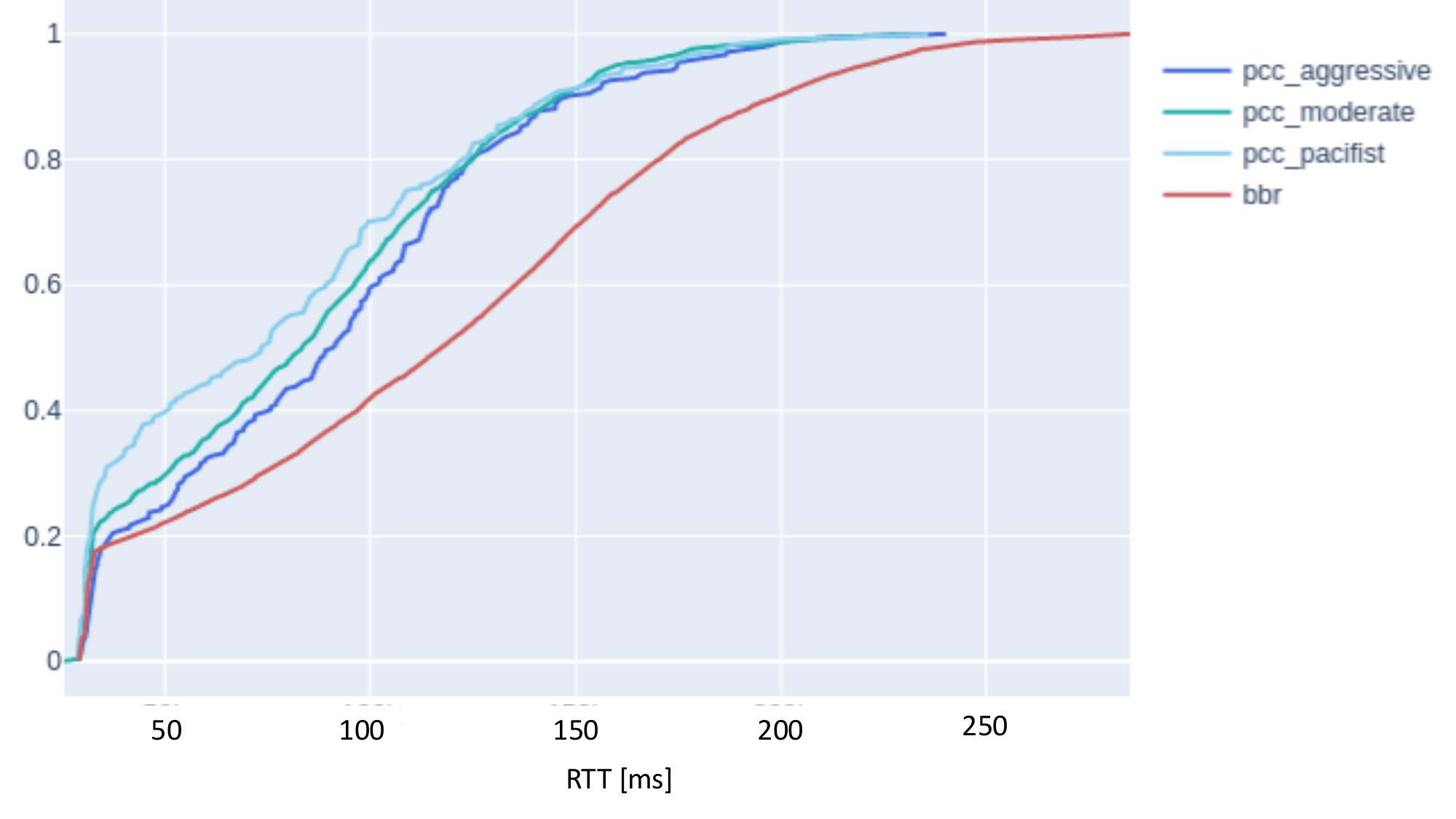} 
        \caption{CDF of average connection RTT}
        \label{fig:meta-connection-RTT}
    \end{subfigure}
    \begin{subfigure}[t]{0.49\textwidth}
        \centering
        \includegraphics[width=\textwidth]{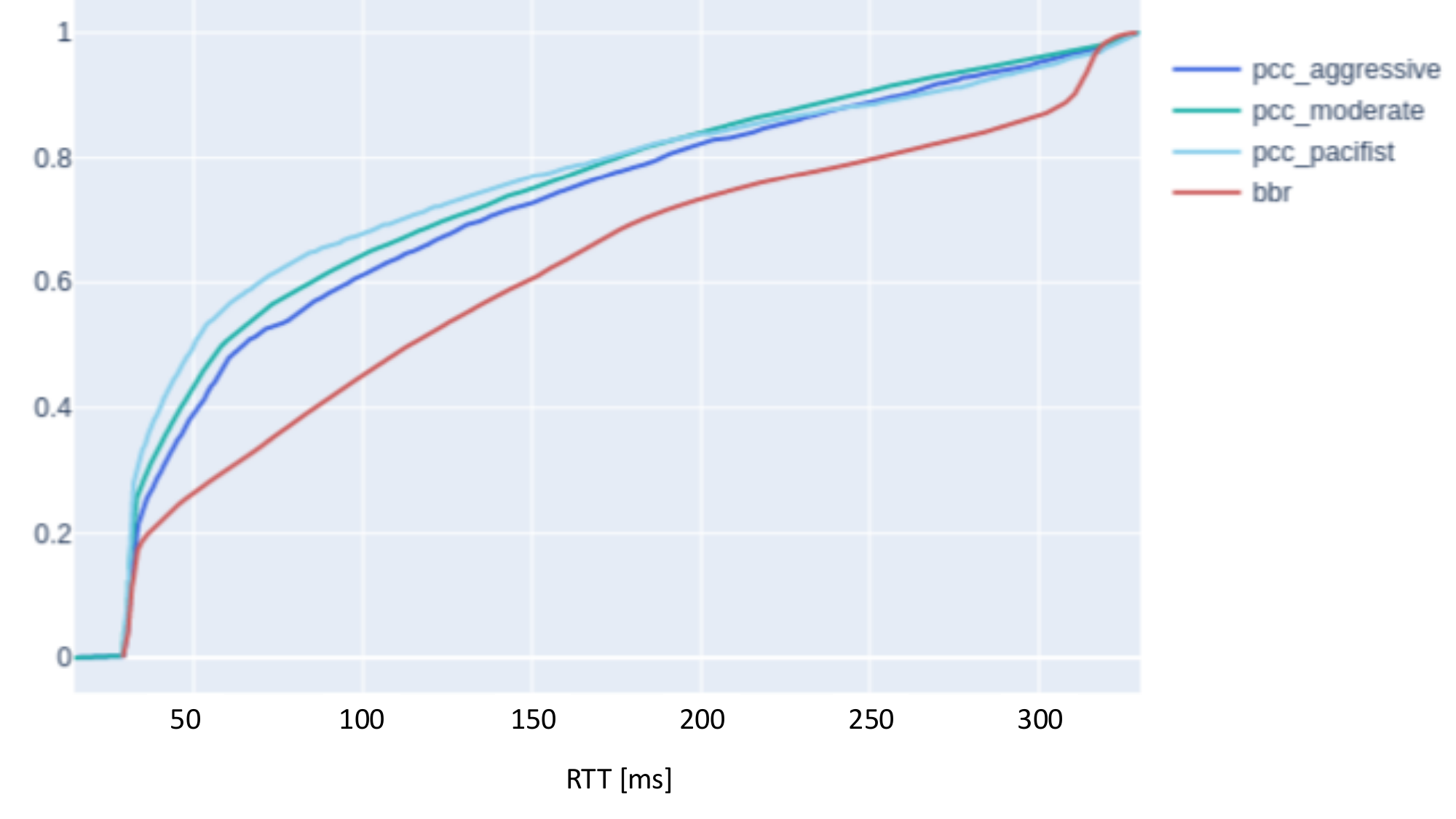} 
        \caption{CDF of RTT by data point}
        \label{fig:meta-dp-RTT}
    \end{subfigure}
    \caption{RTT comparison of BBR and the $3$ learned PCC configurations}
    \label{fig:meta-RTT}
\end{figure*}

\begin{figure*}[h!]
    \centering
    \begin{subfigure}[t]{0.49\textwidth}
        \centering
        \includegraphics[width=\textwidth]{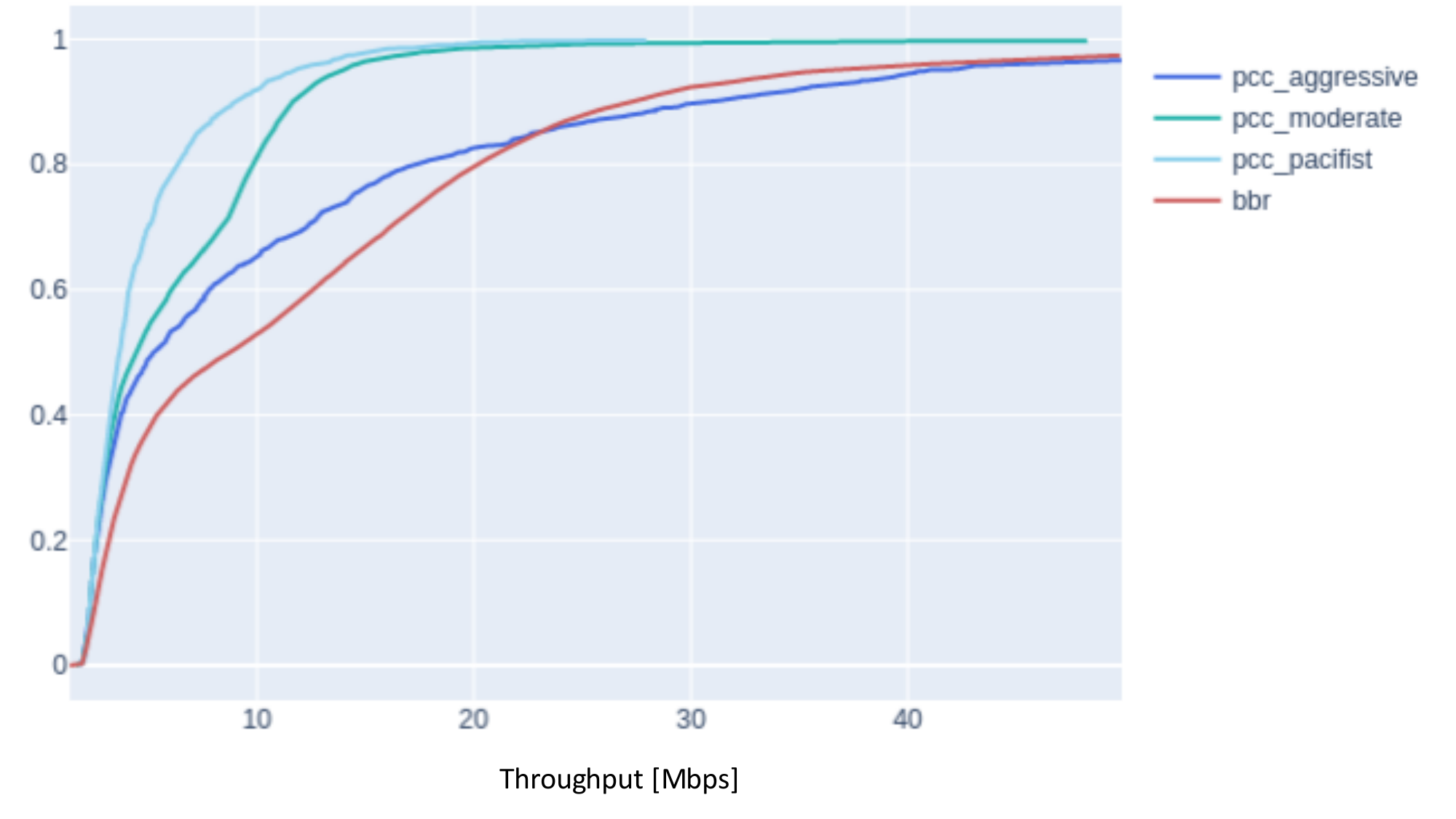} 
        \caption{CDF of throughput}
        \label{fig:meta-tpt-all}
    \end{subfigure}
    \begin{subfigure}[t]{0.49\textwidth}
        \centering
        \includegraphics[width=\textwidth]{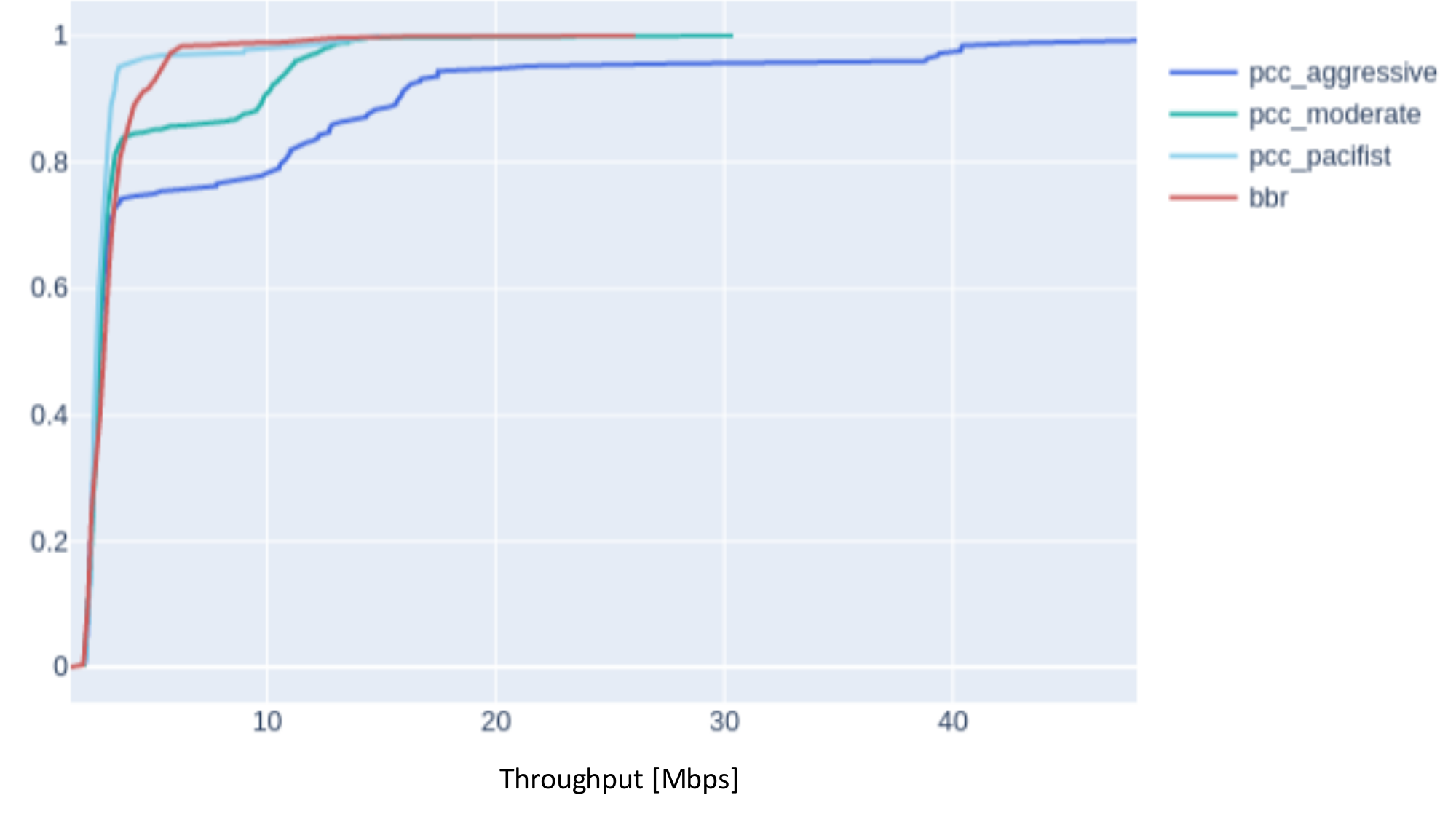} 
        \caption{CDF of throughput for files of size $\geq 600$KB}
        \label{fig:meta-tpt-600}
    \end{subfigure}
    \caption{Throughput comparison of BBR and the $3$ learned PCC configurations}
    \label{fig:meta-tpt}
\end{figure*}

\section{Empirical Results}

We present empirical results from deployments across multiple services (TV streaming, VoD, short-form videos, game downloads) and regions (North and South America, Northeast Asia, the Middle East). We also present results from pilots for connected cars and challenging network conditions (traffic policing). We first detail results for short-form videos from live deployment (\autoref{subsec:SE-asia}) and under traffic policing in an emulated environment (\autoref{subsec:policing}), highlighting key aspects of our customization process and operational lessons, before discussing empirical results for other verticals.

\subsection{Short-Form Videos in Southeast Asia}\label{subsec:SE-asia}

Short-form video delivery (\textit{e.g.}, TikTok, FB Reels) poses significant challenges due to short file sizes, bursty traffic, long silent periods, and reliance on volatile cellular/mobile networks.

\vspace{0.1in}\noindent{\bf Deployment environment.} Our solution was deployed across several countries and benchmarked against BBR by a CDN distributing TikTok content in Southeast Asia. We focus on results from one country, where our solution ran on over $20$ servers in $6$ points of presence (PoPs), serving more than $200$M TikTok videos per server per day. The results, based on A/B testing against BBR, were obtained over $4$ weeks. Both solutions were executed in parallel on separate servers within the same physical location, ensuring identical TikTok request distribution between BBR-controlled and PCC-controlled servers.

\vspace{0.1in}\noindent{\bf Importance of surrogate reward function.} In this multi-CDN scenario, TikTok provided QoE statistics to the CDN, reflecting its competitiveness with other CDNs in that country. However, since TikTok data was available only in daily digests, using it directly for reward computation would slow down decision-making. Instead, our customization engine used a surrogate reward (see \autoref{subsec:reward-engineering}) based on transport-layer data and CDN logs. This surrogate was derived by correlating transport-layer statistics (including percentiles of throughput, RTT, loss rates) and CDN logs (e.g., percentiles of HTTP segment throughput) with TikTok scores, resulting in a linear combination of the most highly correlated statistics (including average HTTP segment throughput and median RTT).

\vspace{0.1in}\noindent{\bf Results for throughput and RTT.} \autoref{fig:TikTok-rewards} presents an A/B comparison to BBR in terms of statistics integrated into our (surrogate) reward function: throughput, as quantified by CDN logs, and (transport-layer) RTT. As seen in the figure, the learned PCC configuration dominates BBR in terms of throughput while simultaneously achieving somewhat lower RTTs.

\vspace{0.1in}\noindent{\bf Implications for TikTok performance scores.} Our surrogate reward function is designed to correlate with TikTok user experience, ensuring that optimizing it also improves QoE. \autoref{fig:TikTok-QoE} shows daily TikTok QoE scores (y-axis) over several months (x-axis), with deployment milestones marked. These scores aggregate many application-layer factors, e.g., video stalls and time-to-first-byte. As our solution was gradually deployed, the CDN's QoE improved relative to the average CDN score in this country, with continued QoE gains even after full adoption due to the ongoing customization by our cloud element.

\vspace{0.1in}\noindent{\bf What did the customization engine learn?} Since CC parameter configuration is independent for different IP subnets, chosen parameter values vary across destinations subnets. An examination of value assignments, however, reveals that, in general, the engine identified key configuration knobs that significantly impacted performance: initial slow start, the latency and loss penalty terms in PCC's utility function, and the loss filter (\autoref{subsec:parametrized-PCC}). This can be interpreted as follows. The initial slow start rate is crucial for short-form videos, which are characterized by short and bursty connections, requiring high bandwidth use without overshooting capacity. The learned configuration was sensitive to latency but robust to loss, as mobile network loss can stem from various factors unrelated to congestion, while packet delays reliably signal congestion in this region, where in-network buffers are deep.

\vspace{0.1in}\noindent{\bf Implications for PCC Vivace.} This deployment, involving short, bursty, app-limited connections over a volatile network, drives many of the modifications to PCC Vivace outlined in \autoref{sec:prototype-to-real}.

\subsection{Short-Form Videos under Policing}\label{subsec:policing}

Traffic policing is implemented by some ISPs to prevent content providers from consuming excessive bandwidth. This can severely impact traffic, causing sudden spikes in loss and packet delays~\cite{policing}. One common method, the token bucket, generates tokens at a set rate and throttles the connection when tokens are depleted. This creates significant challenges for congestion control protocols, as the network conditions can abruptly shift from optimal to poor without any prior congestion signals.

\vspace{0.1in}\noindent{\bf Experimental Setup.} In a pilot with a large content provider, our solution was compared to BBR in an emulated network environment simulating harsh real-world conditions induced by traffic policing. A traffic generator developed by the content provider mimicked short-form video traffic based on file time and size distribution. The evaluation focused on transport-layer metrics (throughput, loss, RTT), making a surrogate reward unnecessary.

\begin{figure*}[h!]
    \begin{subfigure}{0.49\textwidth}
        \includegraphics[width=\textwidth]{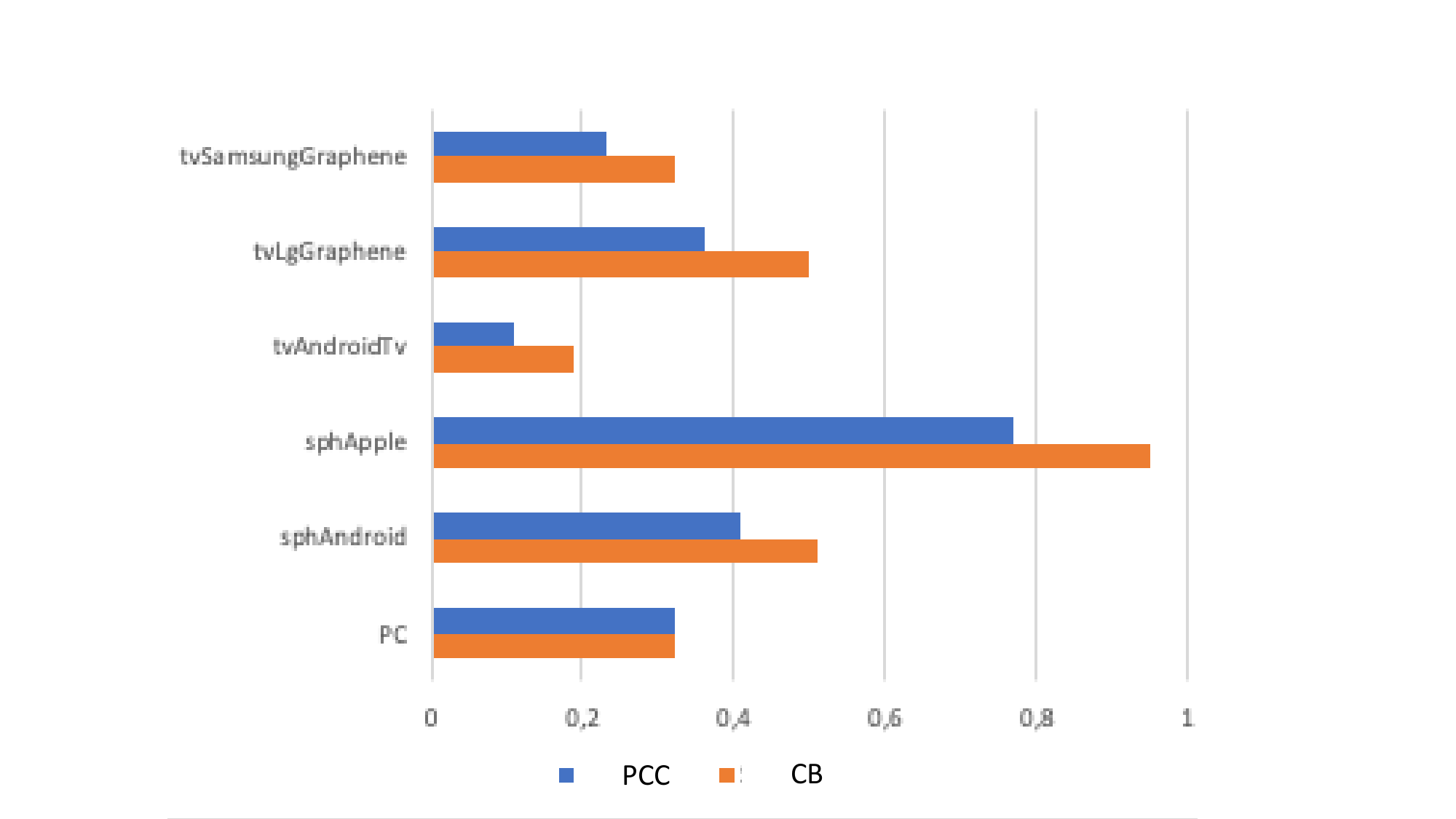} 
        \caption{LATAM1 rebuffering ratios}
        \label{fig:LATAM1-rebuffer}
    \end{subfigure}
    \begin{subfigure}{0.49\textwidth}
        \includegraphics[width=\textwidth]{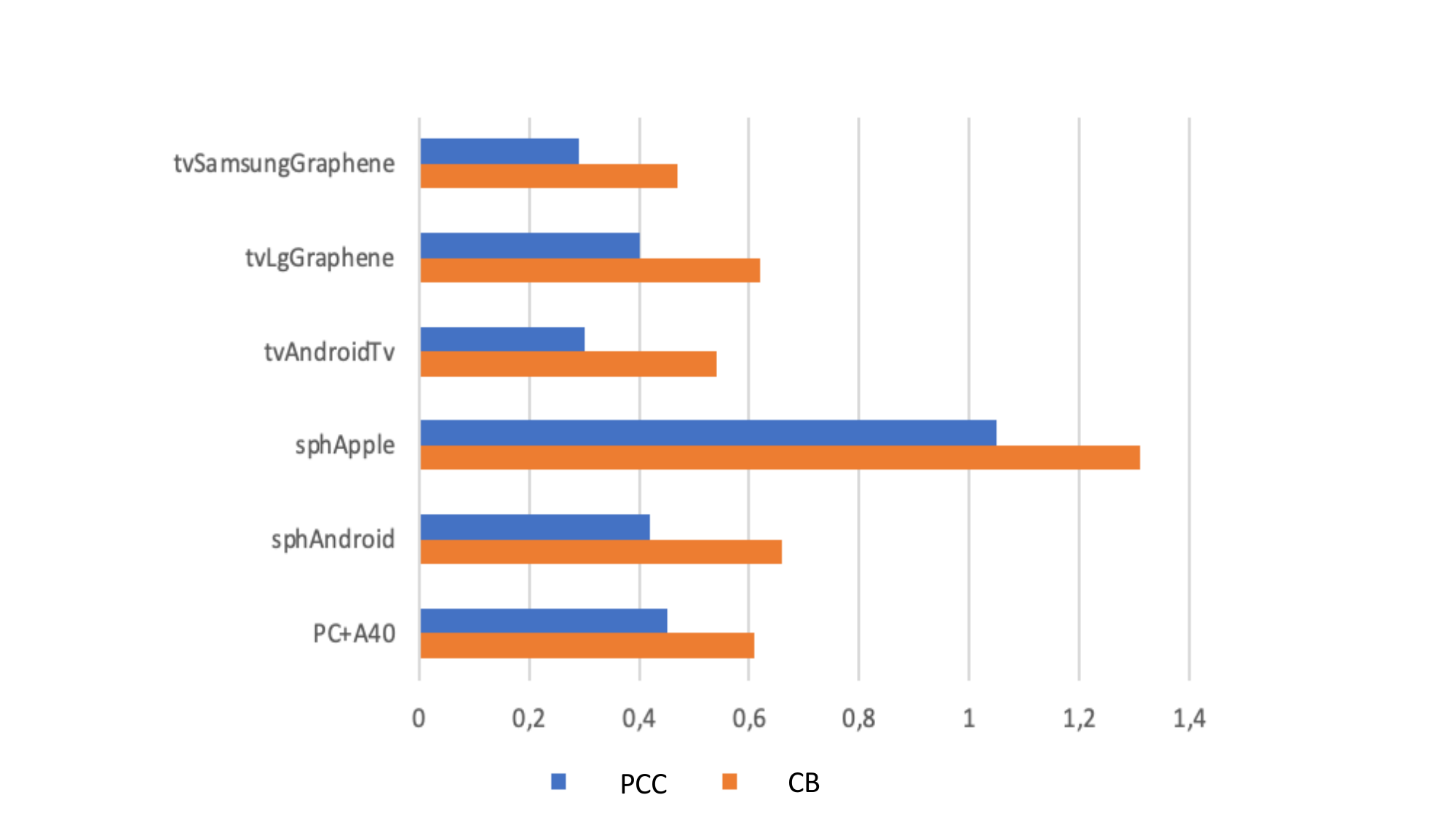} 
        \caption{LATAM2 rebuffering ratios}
        \label{fig:LATAM2-rebuffers}
    \end{subfigure}
    \caption{Rebuffering ratios for LATAM1 and LATAM2, broken down by device type}
    \label{fig:LATAM-rebuffer}
\end{figure*}

\begin{figure*}[h!]
    \centering
    \begin{subfigure}{0.49\textwidth}
        \includegraphics[width=\textwidth]{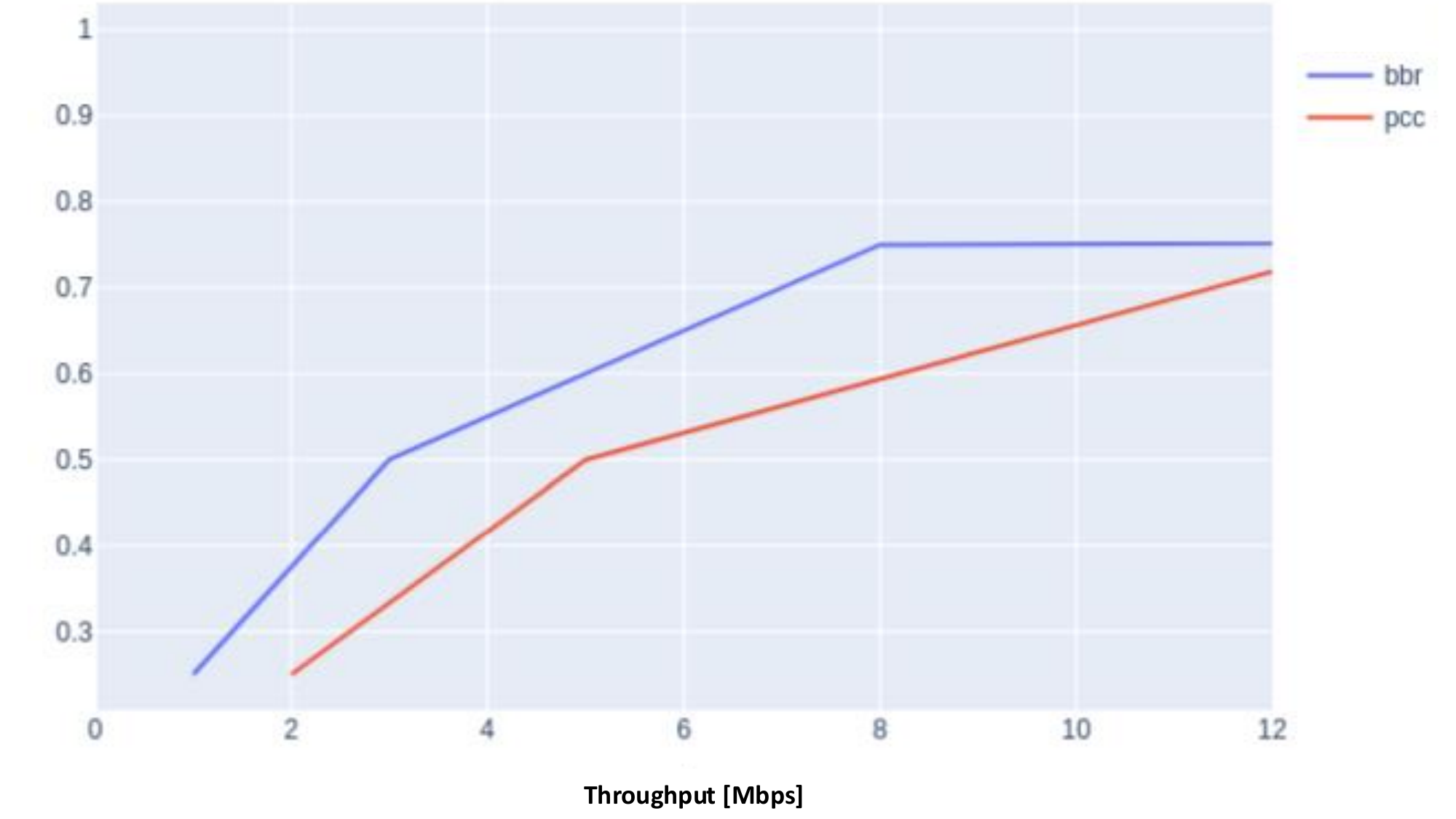} 
        \caption{CDF of game download throughput}
        \label{fig:Lumen-tpt}
    \end{subfigure}
    \begin{subfigure}{0.49\textwidth}
        \includegraphics[width=\textwidth]{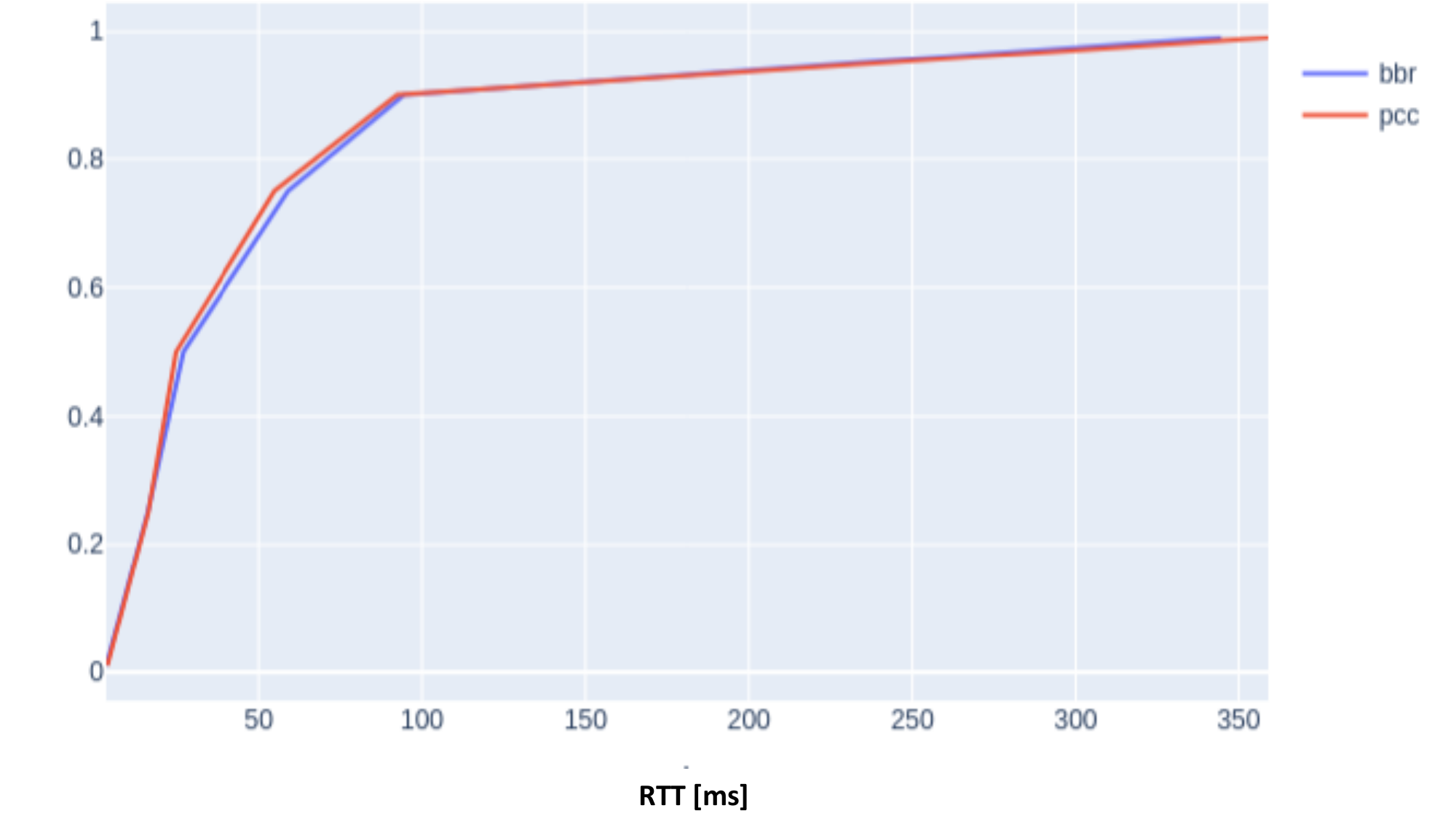} 
        \caption{CDF of RTT}
        \label{fig:Lumen-RTT}
    \end{subfigure}
    \caption{Results for game downloads}
    \label{fig:Lumen}
\end{figure*}

\vspace{0.1in}\noindent{\bf $3$ learned PCC configurations.} To demonstrate our solution's flexibility, we formulated three reward functions of the form $\alpha T - \beta D - \gamma L$, where $T$, $D$, and $L$ represent average throughput, $75$th percentile RTT, and average packet loss, respectively. The reward functions varied in the weights assigned to these factors: ``\textit{Pacifist}'' prioritized RTT and loss minimization, ``\textit{Aggressive}'' prioritized throughput, and ``\textit{Moderate}'' lies between the two. The PCC configuration with respect to each of these reward functions was learned over $12$ hours and evaluated over $4$ hours.

\vspace{0.1in}\noindent{\bf Results.} \autoref{fig:meta-loss} compares loss rates for BBR and the PCC configurations learned with three different rewards. \autoref{fig:meta-RTT} shows RTT comparisons, with \autoref{fig:meta-connection-RTT} and \autoref{fig:meta-dp-RTT} plotting CDFs for average RTT per connection and $1$-second intervals, respectively. As expected, the PCC configurations are ranked from Pacifist (low) to Aggressive (high) in terms of loss and RTT. Pacifist (and even Moderate) significantly reduce loss compared to BBR, which experiences high packet loss. All three PCC configurations outperform BBR in RTT.

\autoref{fig:meta-tpt-all} compares throughput, with Pacifist, Moderate, and Aggressive ordered from low to high, as expected. BBR outperforms Pacifist and Moderate in throughput, while Aggressive only surpasses BBR at higher percentiles. A closer look at the results, however, reveals a more nuanced picture: the bigger the file size, the better PCC performs. In particular, as shown in \autoref{fig:meta-tpt-600} for files over $600$KB, which make up over $40\%$ of transmitted data, even Pacifist matches BBR's throughput across most percentiles, with Moderate and Aggressive outperforming BBR at higher percentiles. Since larger files correspond to higher video qualities, achieving higher throughputs for small files and lower throughputs for the larger files, as with BBR, is undesirable.

\vspace{0.1in}\noindent{\bf What did the customization engine learn?} Examining the engine's value assignments indicates that the configuration engine learned an appropriate initial slow start rate, to cap the rate at which connections leave slow start, and to become highly sensitive to loss. This behavior contrasts with learned configurations in \autoref{subsec:SE-asia}, and can be explained as follows. In traffic-policed environments, network conditions remain good until tokens are exhausted, followed by sharp increases in latency and loss. Capping the maximum rate post-slow-start and aggressively reacting to loss or latency spikes helps mitigate these issues.

\vspace{0.1in}\noindent{\bf Implications for PCC Vivace.} In this environment, the sender must react quickly to congestion, which often appears only after the rate exceeds available bandwidth by a lot. The modifications to PCC that accelerate decision-making (see \autoref{sec:prototype-to-real}) and, in particular, the introduction of emergency breaks, were crucial.

\subsection{TV Streaming in LATAM}

We present results from production deployment at a global content provider's CDN in two Latin American countries (LATAM1 and LATAM2), which have different user device and access network distributions (see \autoref{fig:LATAM-device} and \autoref{fig:LATAM-access}). The benchmark in both countries is TCP Cubic on Linux. The results presented for each country are based on A/B testing within the same physical point of presence over the course of a month, contrasting the performance of one server running our PCC implementation and another running Cubic, with both servers supporting identical video client distribution.

In this deployment, our cloud element had access to QoE data from video clients, allowing the reward function to incorporate QoE information without needing a surrogate. Since rebuffering significantly impacts user engagement, the customization engine prioritized minimizing rebuffering times, a key concern for TV streaming providers. Both countries experienced fairly frequent and lengthy rebuffering events, making this focus crucial. The reward function was $c_B B - c_R R$, where $B$ is the average video bitrate and $R$ is the average rebuffering ratio, with coefficients $c_B$ and $c_R$ set to heavily penalize rebuffering.

The learned PCC configurations significantly reduced rebuffering ratio and time. In LATAM1, the average rebuffering ratio and time decreased by $24\%$ and $36\%$, respectively, while in LATAM2, they dropped by $29\%$ and $31\%$. The largest improvements were seen in the $90$th percentile rebuffering time, with reductions of $29\%$ in LATAM1 and $32\%$ in LATAM2. \autoref{fig:LATAM-rebuffer} shows the average rebuffering ratios for PCC (our edge) and Cubic (CB) across device types, with our solution reducing rebuffering ratios for nearly all device types. Our also solution increased video bitrates at lower percentiles ($19\%$ and $144\%$ at the $25$th percentile for LATAM1 and LATAM2, respectively) without affecting higher percentiles. In LATAM2, the number of sessions watched at most in $270$p resolution decreased by $31\%$, while those viewed at least in $720$p rose by $22\%$.

\subsection{Game Downloads in North America}

We next discuss deployment by a large CDN in North America for the use-case of game downloads (Sony PlayStation). The results are based on A/B testing against BBR (in Linux), and contrast the performance of one server running PCC and one server running BBR over the course of $3$ weeks. The target metric was reducing median game download times, and so our reward function explicitly targeted this metric. Our solution reduced median game download time by $>50\%$. This 
was accomplished by learning PCC configurations that substantially improve the throughput of downloading game ``chunks'', especially at the lower percentiles (\autoref{fig:Lumen-tpt}), while not harming network latency (\autoref{fig:Lumen-RTT}). (Since the throughput curves largely coincide from the $75$th percentile onwards, and since bad QoE is associated with lower throughputs, we cap the x-axis at $12$Mbps for clarity.)

\begin{figure*}[h!]
    \centering
    \begin{subfigure}{0.49\textwidth}
    \includegraphics[width=\textwidth]{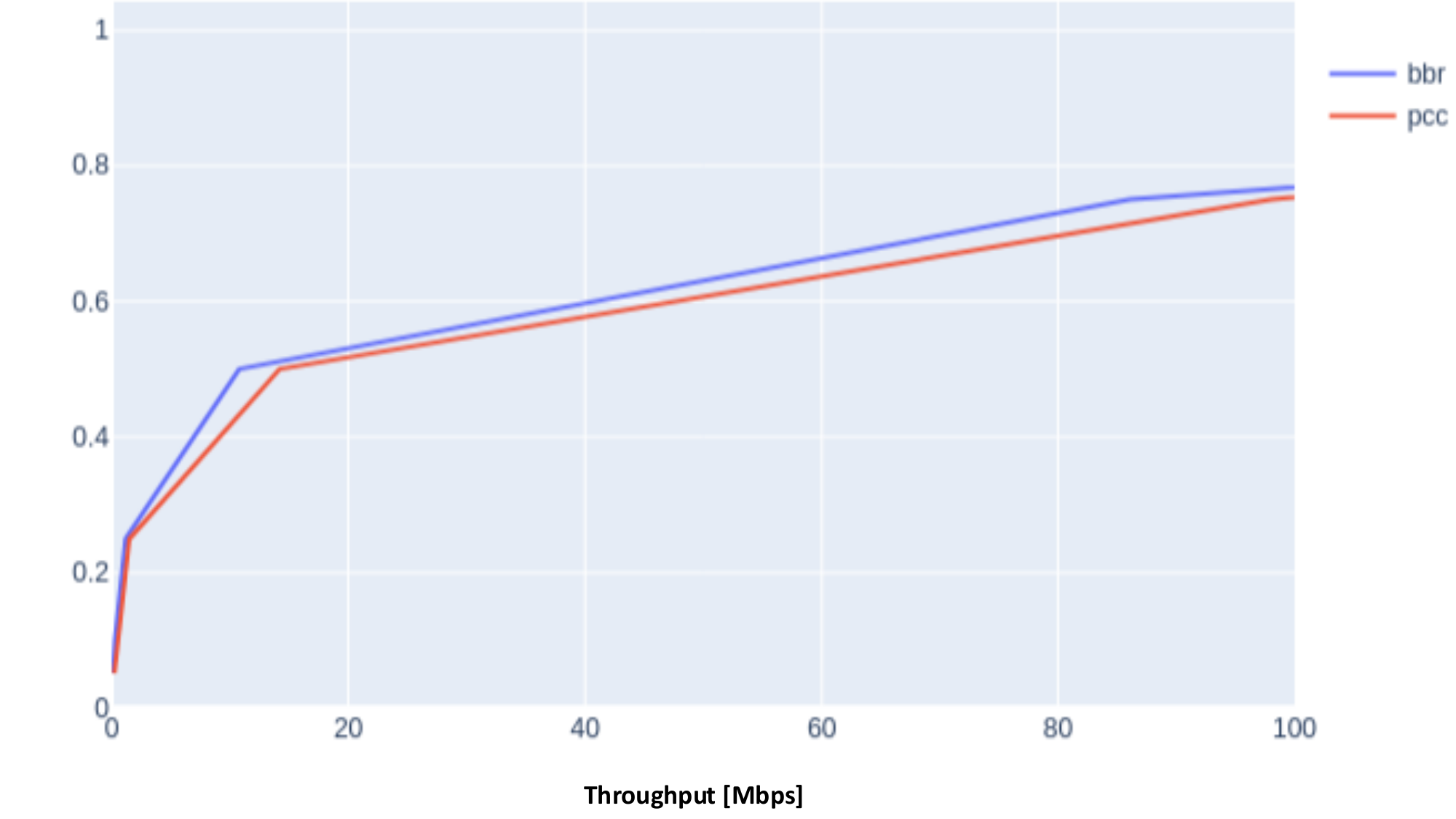} 
        \caption{CDF of throughput}
        \label{fig:VoD-tpt}
    \end{subfigure}
    \begin{subfigure}{0.49\textwidth}
        \includegraphics[width=\textwidth]{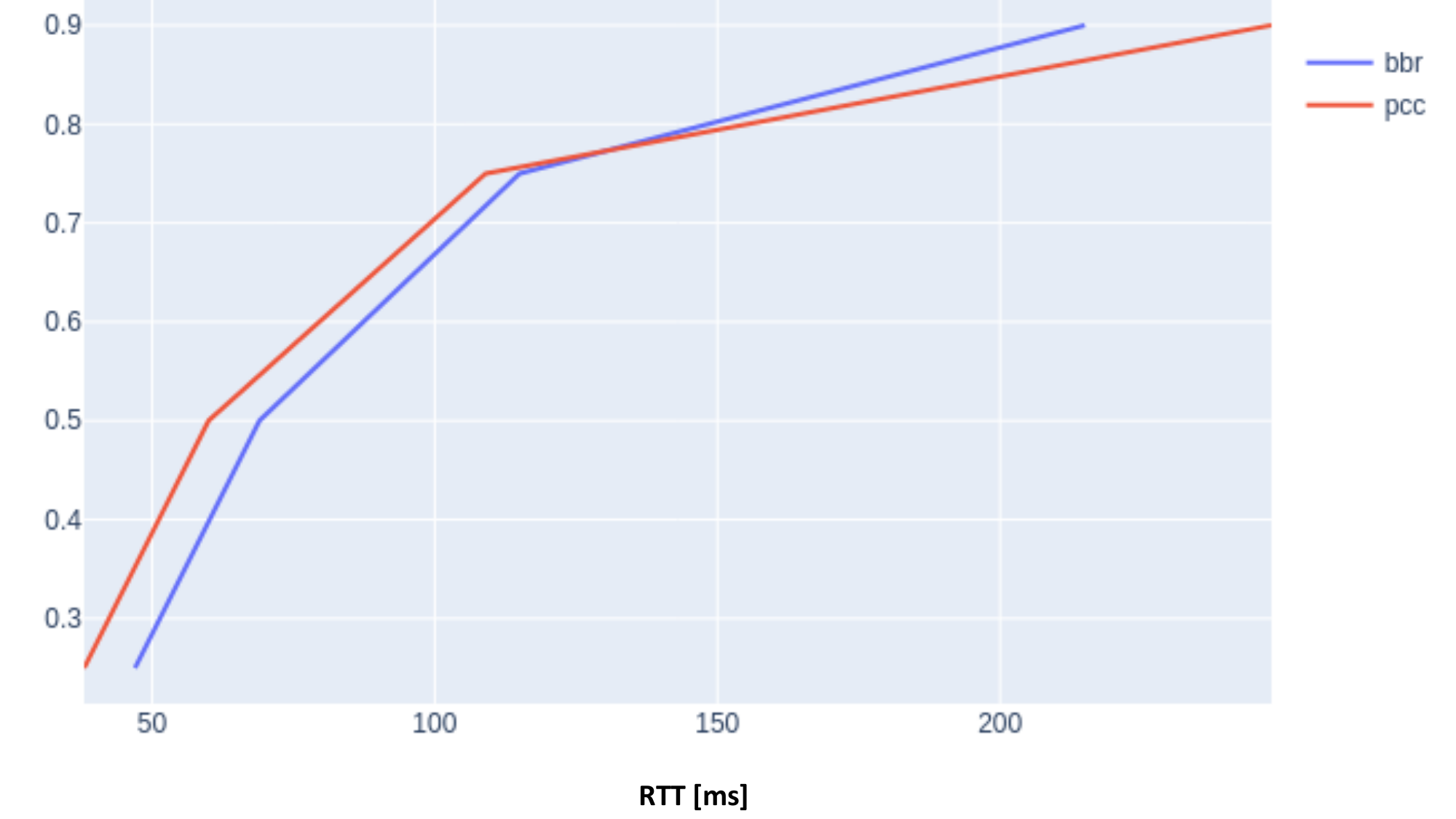} 
        \caption{CDF of RTT}
        \label{fig:VoD-RTT}
    \end{subfigure}
    \caption{Results for VoD}
    \label{fig:VoD}
\end{figure*}

\begin{figure*}[h!]
    \centering
    \begin{subfigure}[t]{0.49\textwidth}
        \centering
        \includegraphics[width=\textwidth]{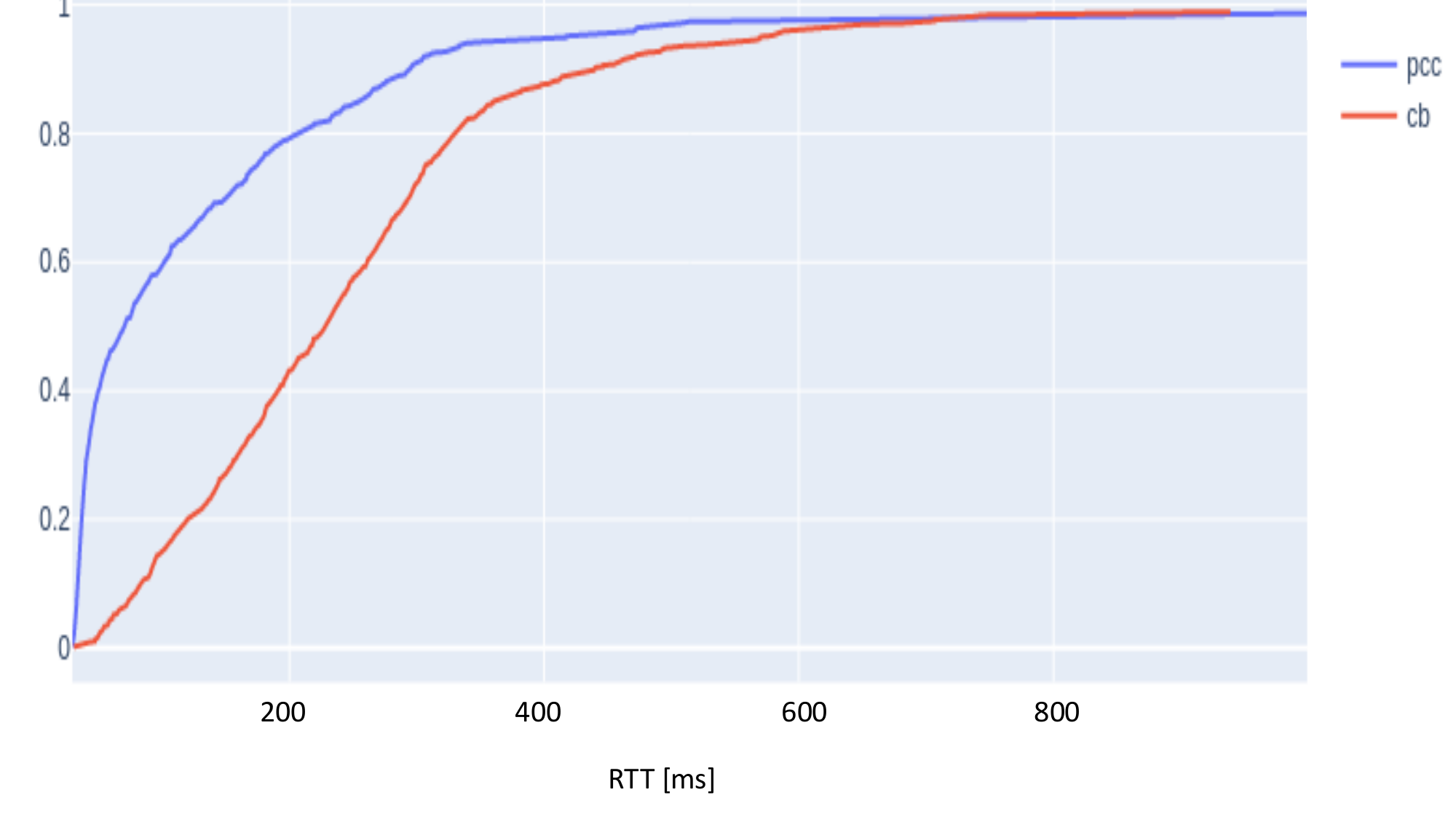} 
        \caption{CDF of RTT}
        \label{fig:cars-RTT}
    \end{subfigure}
    \begin{subfigure}[t]{0.49\textwidth}
        \centering
        \includegraphics[width=\textwidth]{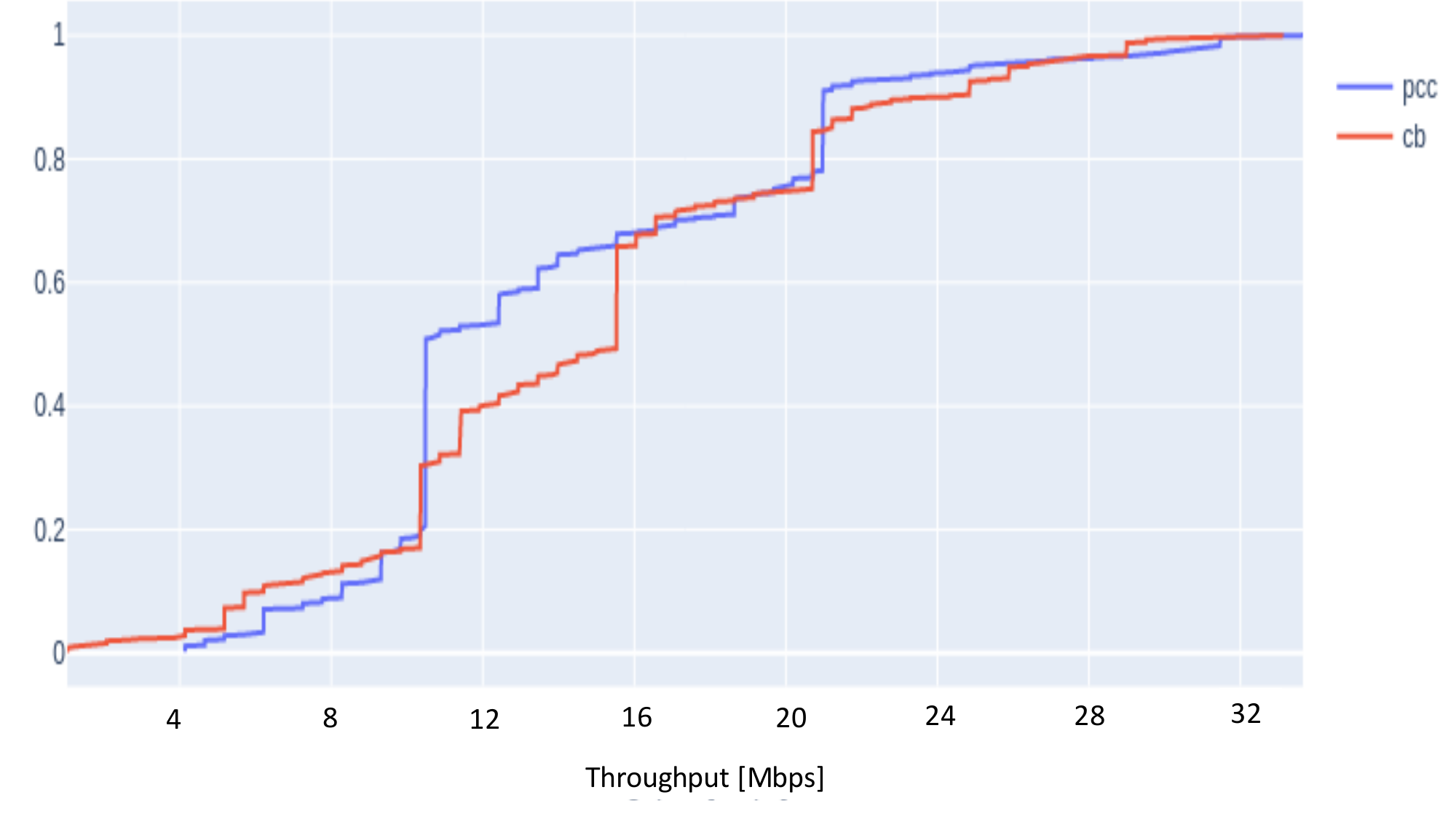} 
        \caption{CDF of throughput}
        \label{fig:cars-tpt}
    \end{subfigure}
    \caption{Results for connected car experiment}
    \label{fig:connected-cars}
\end{figure*}

\subsection{VoD in the Middle East}

The next use-case involves deployment in production for video-on-demand by a large content distributor in the Middle East. The benchmark here was BBR. A/B testing was performed over the course of a month with $6$ servers running each protocol. A surrogate reward function that targeted a tradeoff between throughput and RTT was used. Learned PCC configurations improved average rebuffering ratio by $32\%$, and reduced the number of sessions with extreme rebuffering events, defined as sessions with rebuffering ratio of at least $2\%$, by $28\%$. This was achieved in parallel to significantly improving video download throughput, which manifested in both higher average video bitrate and faster video start-time ($17\%$ reduction; from $2.6$s to $2.2$s).

VoD applications are less sensitive to latency than live streaming, as the playback buffer can absorb network delays.  VoD performance is more affected by network throughput, which influences video quality, start time, and rebuffering. \autoref{fig:VoD-tpt} and \autoref{fig:VoD-RTT} show the CDFs of HTTP throughput and connection RTT from the learned PCC configuration. Since throughput curves converge after the $75$th percentile and low throughput correlates with poor QoE, the x-axis is capped at $100$Mbps for clarity. Throughput and latency typically trade off, as increasing throughput fills network buffers, raising packet delays. The customization engine increased throughput (e.g., $30\%$ higher median). Interestingly, \autoref{fig:VoD-RTT} reveals a mixed picture: lower RTTs than BBR at most percentiles ($0-75$th) and higher RTTs at the higher percentiles. A closer look at the data reveals the reason: for bigger HTTP-responses, which represent higher quality video chunks, the learned PCC configuration achieves higher throughputs at the expense of increasing RTTs.

\subsection{Connected Cars}

Delivering content to and from vehicles is challenging due to variability in network bandwidth, latency, intermittent congestion, and handovers causing rate drops and packet loss. We report results from a live pilot with a major car manufacturer, where our edge solution was installed in a moving vehicle driving in an LTE network environment. Files of various sizes were continuously uploaded to the nearest public cloud (within the same country in Asia), with comparisons made to the car's existing TCP Cubic mechanism. The experiment consisted of $5$ $30$-minute-long drives in both highways and city roads, switching between PCC and BBR every $30$ seconds. RTT and throughput measurements were collected by our edge element and aggregated every second.

The objective was to reduce latency between the car and the public cloud, which is crucial for real-time data delivery like live video from car cameras. To achieve this, our reward function incorporated both throughput and latency terms, with a strong penalty for increased latency.\autoref{fig:cars-RTT} and \autoref{fig:cars-tpt} show results for RTT and throughput. Due to deep in-network buffers, RTT for TCP Cubic can be extremely high. With our solution, median RTT improved by $3$x (from $228$ms to $70$ms), and average RTT improved by $1.7$x (from 256ms to 145ms), with improvements across all percentiles. However, improving RTT resulted in a $5\%$ average throughput decrease ($1.85$MBps to $1.76$MBps) and a $32\%$ median decrease ($1.94$MBps to $1.31$MBps). Different latency-throughput tradeoffs can be targeted by adjusting the reward function weights.

\section{Discussion}\label{sec:discussion}

The success of our solution in improving different notions of performance across different use-cases evidences the promise of automated customization of congestion control. We believe, however, that we still have a long way to go to fully tap the potential of this approach. We mention a few interesting directions below. 

\vspace{0.1in}\noindent{\bf Leveraging unsupervised learning to improve context and configuration-aggregate generation.} Various components of our solution rely on clustering similar elements, such as generating contexts for the customization algorithm and determining configuration aggregates. While simple heuristics perform well in practice (\autoref{subsec:customization-detailed}), applying unsupervised learning algorithms to optimize clustering could enhance performance.

\vspace{0.1in}\noindent{\bf Investigating alternative learning-based approaches for CC customization.} Our customization algorithm follows a contextual continuum-armed bandits approach. An intriguing direction is exploring other methods, such as deep reinforcement learning (DRL), which could also potentially eliminate the need for clustering (see above). However, more complex approaches may introduce challenges, such as difficulties in reasoning about and debugging decisions, and the need for more data samples to learn effective PCC configurations.

\vspace{0.1in}\noindent{\bf Cross-aggregate learning.} Currently, our algorithm customizes each configuration aggregate independently. However, insights gained from one environment or aggregate could be valuable for similar ones, akin to transfer learning.

\section{Related Work}


\vspace{0.1in}\noindent{\bf Learning-based CC.} Various approaches to learning-based CC have been proposed. Unlike our approach, \textit{all} of these approaches are application-agnostic. Remy~\cite{tcp_ex_machina} customizes CC to a network environment using as input \textit{human-provided assumptions} regarding the network. Past academic work on PCC~\cite{pcc,vivace,pcc_proteus} employs online learning for CC, but still constitutes a one-size-fits-all solution. Aurora~\cite{aurora} and Orca~\cite{orca} use deep RL to learn one-size-fits all CC policies from \emph{synthetic} network environments.

\vspace{0.1in}\noindent{\bf Customizing CC configurations.} The two following academic studies propose automatically configuring CC options. Configanator~\cite{configanator} explores data-driven customization of web serving protocol stack configurations, including congestion control parameters. Importantly, configanator's analyses are focused on a specific, and rather simple, target application-layer metric (page load times) and involve mostly discrete (and often boolean) parameter value domains. Floo~\cite{Floo} utilizes reinforcement learning to dynamically choose between CC protocols so as to optimize request completion times. This, also, entails a discrete (fairly small) set of congestion logic options and a rather simple target performance metric.

\vspace{0.1in}\noindent{\bf Customizing ABR to specific network environment.} \cite{puffer} shows how an adaptive video bitrate (ABR) algorithm can be automatically adapted to its deployment environment by learning in situ. ABR decisions of this deployment-tailored algorithm are still subject to the performance of the underlying (one-size-fits-all) CC protocol.


\section{Conclusion}

We have presented lessons from the design and deployment of a holistic system for automatically customizing congestion control in the wild. Our empirical experience from the deployment of our solution across multiple use-cases, geographical regions, and networks, suggests that customizing congestion control can substantially improve QoE.

\bibliographystyle{plain}
\bibliography{b2.bib}
\newpage

\appendix
\begin{figure*}[h!]
    \centering
    \begin{subfigure}[t]{0.49\textwidth}
        \centering
        \includegraphics[width=\textwidth]{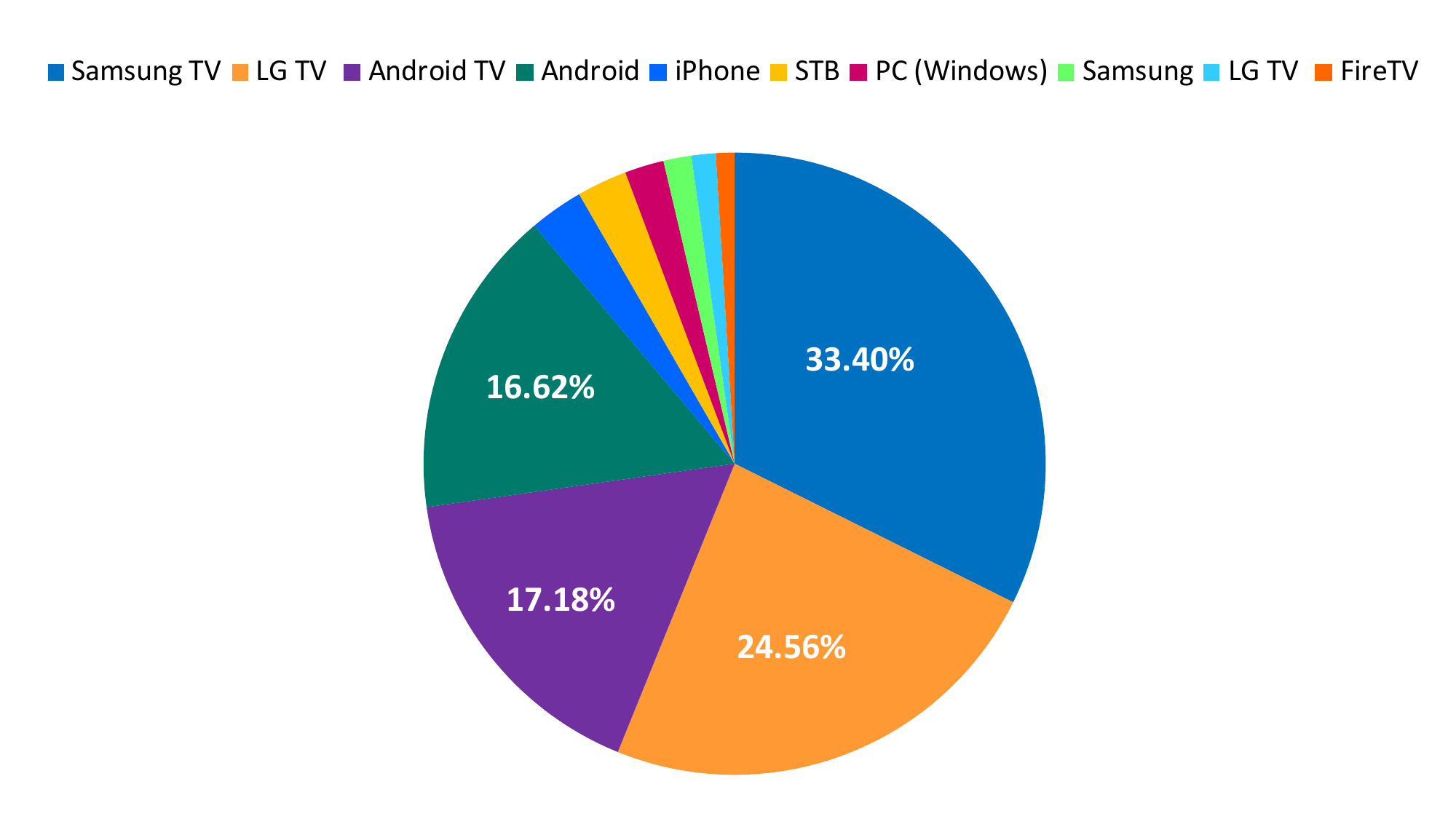} 
        \caption{LATAM1 device distribution}
        \label{fig:LATAM1-device}
    \end{subfigure}
    \begin{subfigure}[t]{0.49\textwidth}
        \centering
        \includegraphics[width=\textwidth]{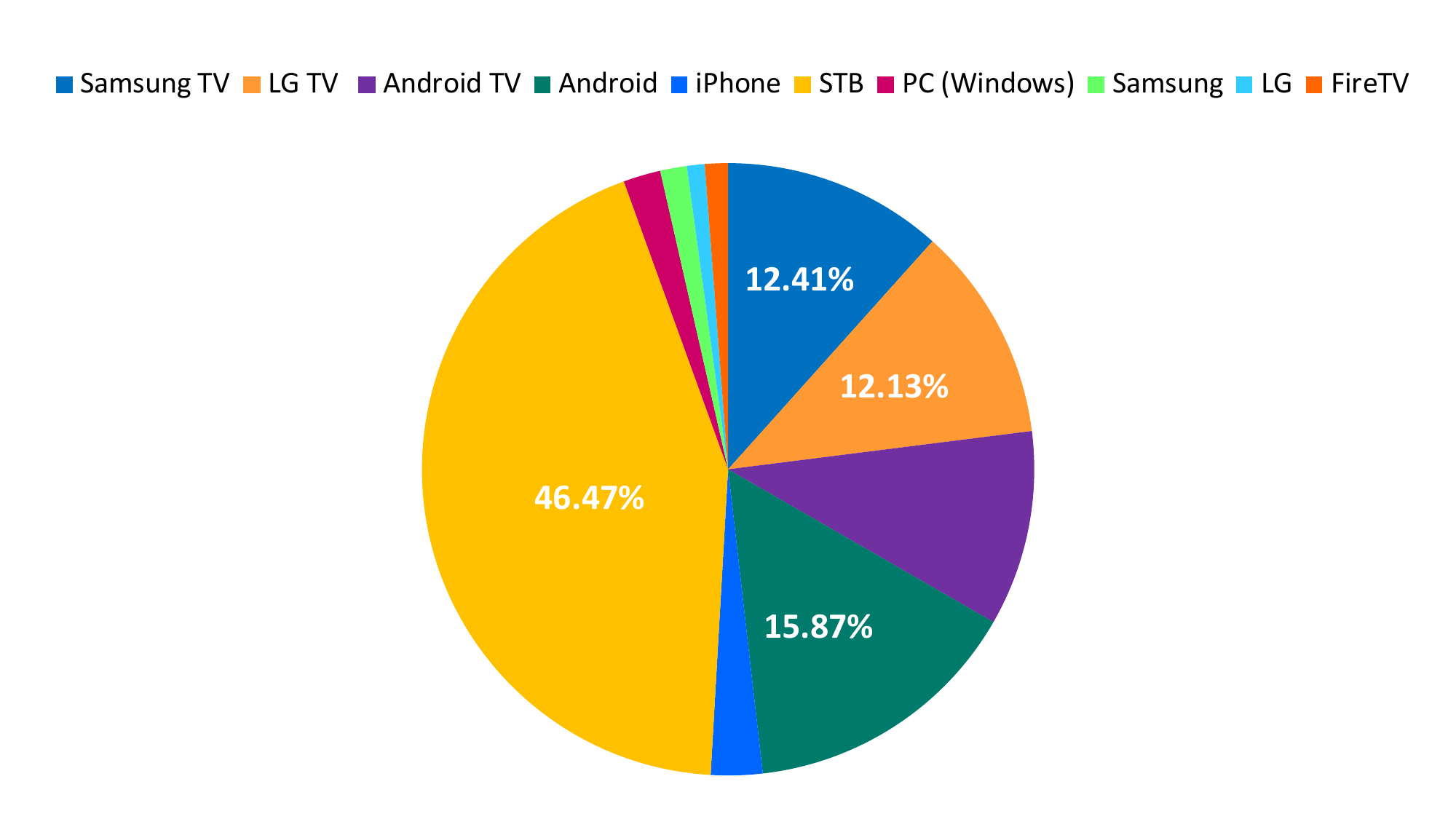} 
        \caption{LATAM2 device distribution}
        \label{fig:LATAM2-device}
    \end{subfigure}
    \caption{Device distribution}
    \label{fig:LATAM-device}
\end{figure*}

\begin{figure*}[h!]
    \centering
    \begin{subfigure}{0.49\textwidth}
        \centering
        \includegraphics[width=\textwidth]{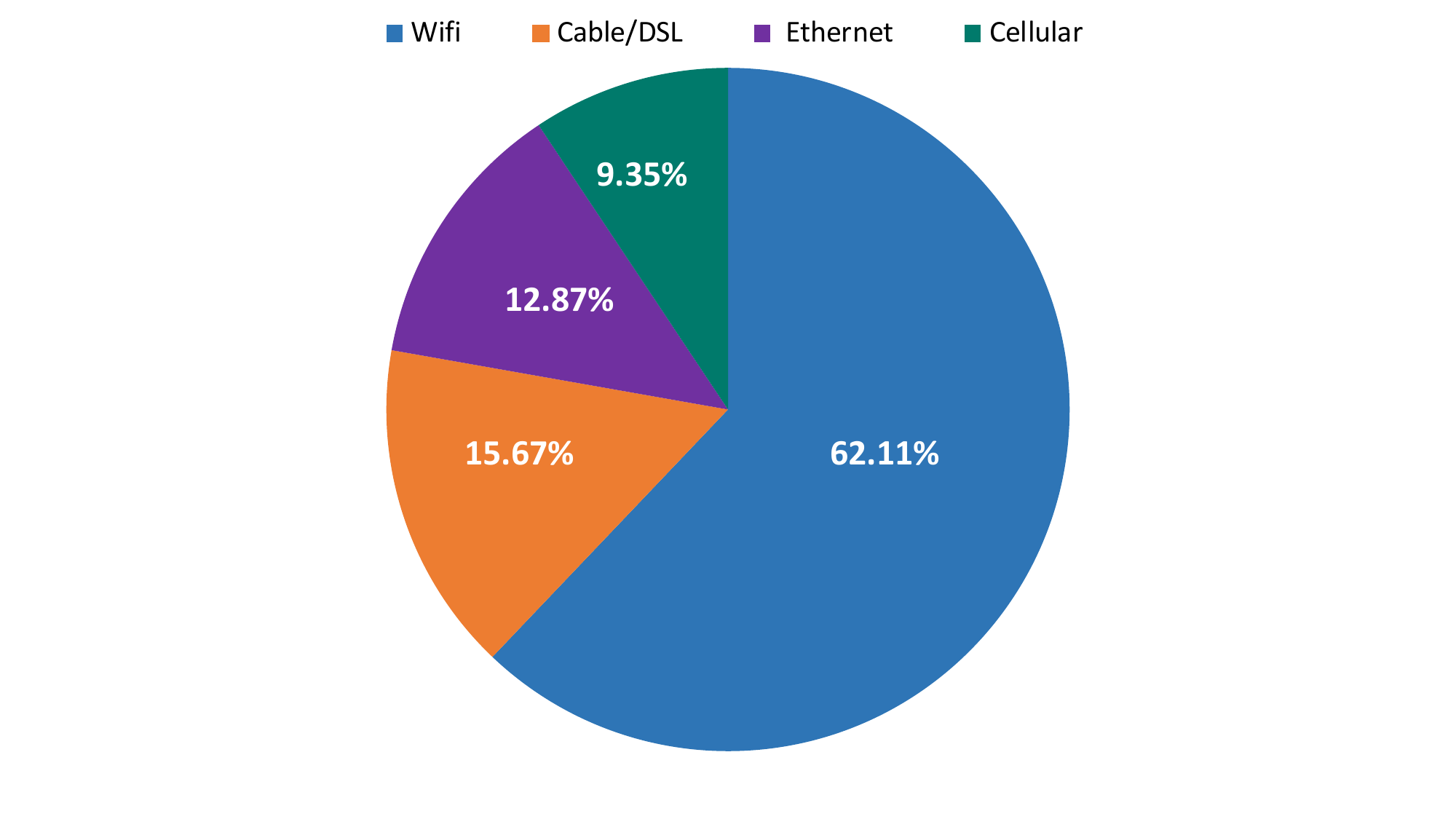} 
        \caption{LATAM1 access distribution}
        \label{fig:LATAM1-access}
    \end{subfigure}
    \begin{subfigure}{0.49\textwidth}
        \centering
        \includegraphics[width=\textwidth]{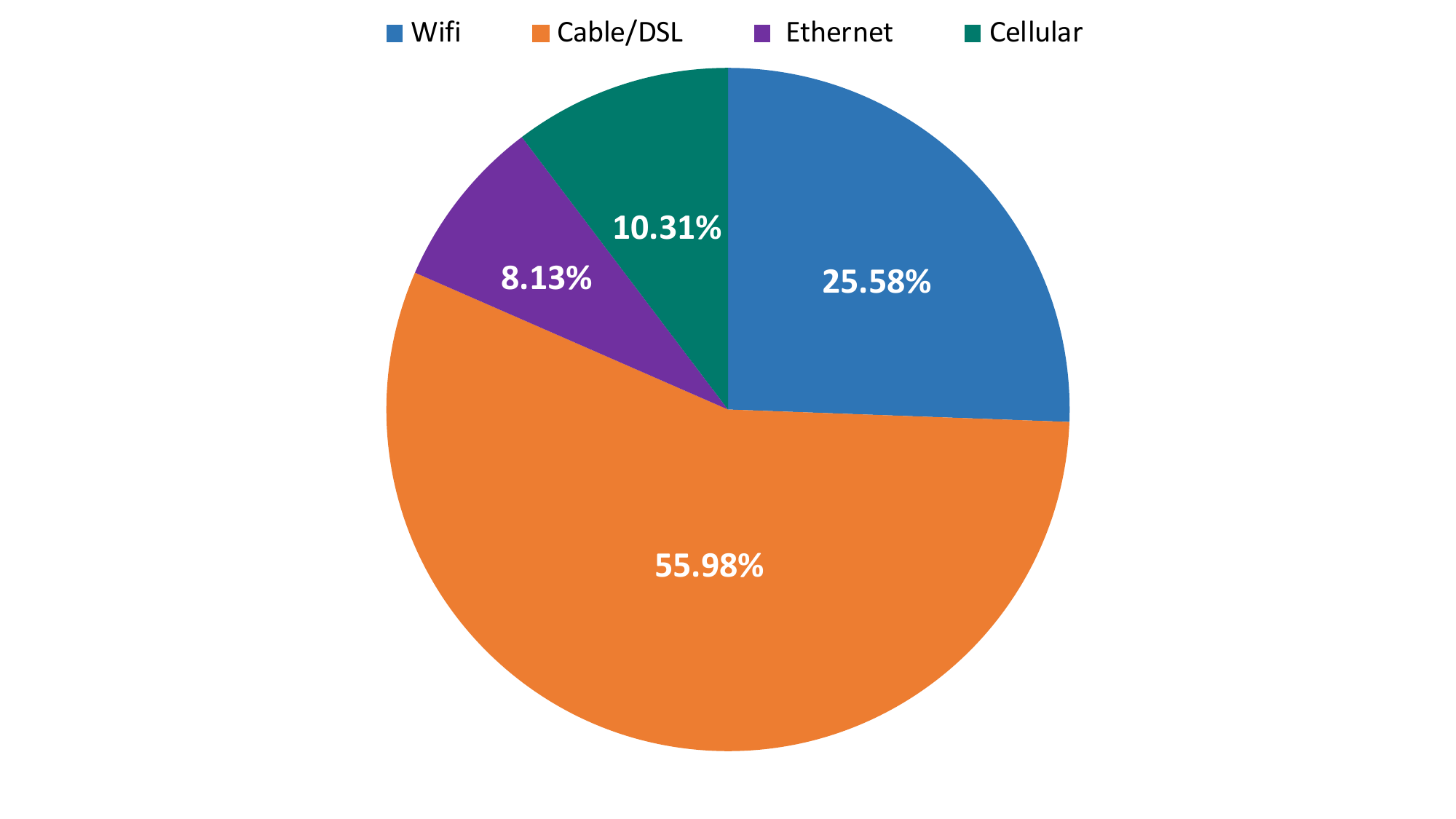} 
        \caption{LATAM2 access distribution}
        \label{fig:LATAM2-access}
    \end{subfigure}
    \caption{Access distribution}
    \label{fig:LATAM-access}
\end{figure*}

\end{document}